\documentclass[conference,compsoc]{IEEEtran}
\IEEEoverridecommandlockouts 
\usepackage[disable]{todonotes}

\usepackage{amsmath,amssymb,amsthm}
\usepackage{booktabs}
\usepackage{multirow}
\usepackage{tabularx}
\usepackage{graphicx}
\usepackage{tikz}
\usetikzlibrary{arrows.meta}
\usepackage{listings}
\usepackage{enumitem}
\usepackage{needspace}
\usepackage{float}
\usepackage{caption}
\usepackage{subcaption}
\usepackage{xspace}
\usepackage{array}
\usepackage{algorithm}       
\usepackage{algpseudocode}   
\usepackage{xcolor}        
\usepackage{bm}

\definecolor{asmBg}{gray}{0.97}
\definecolor{asmGray}{gray}{0.45}
\definecolor{asmGreen}{rgb}{0.00,0.45,0.30}
\definecolor{asmBlue}{rgb}{0.10,0.25,0.65}
\definecolor{asmRed}{rgb}{0.70,0.10,0.15}
\definecolor{asmPurple}{rgb}{0.45,0.15,0.60}

\lstdefinestyle{gadget}{
  basicstyle=\ttfamily\scriptsize,
  numbers=left, numberstyle=\tiny, stepnumber=1, numbersep=5pt,
  frame=single, breaklines=true, captionpos=b,
  morecomment=[l]{//},
  morecomment=[s]{/*}{*/},
  keywordstyle=\bfseries,
  alsoletter={.},
  morekeywords={adr,str,dsb,dc,civac,add,bl,ldp,ldr,ret,mov,subs,umov,cbz,
                madd,mul,eor,nop,b.ne,b.eq},
}

\lstdefinestyle{aarch64spectre}{
  basicstyle=\ttfamily\footnotesize,
  backgroundcolor=\color{asmBg},
  frame=single,
  rulecolor=\color{asmGray!35},
  numbers=left,
  numberstyle=\scriptsize\color{asmGray},
  numbersep=7pt,
  xleftmargin=2em,
  framexleftmargin=1.6em,
  columns=fullflexible,
  keepspaces=true,
  showstringspaces=false,
  breaklines=true,
  captionpos=b,
  morecomment=[l]{//},
  morecomment=[s]{/*}{*/},
  commentstyle=\color{asmGreen}\itshape,
  keywordstyle=\color{asmBlue}\bfseries,
  morekeywords={adr,str,dsb,dc,civac,add,bl,ldp,ldr,ret,mov,subs,umov},
  emph={[1]anchor,ak},
  emphstyle={[1]\color{asmPurple}\bfseries},
}

\usepackage{letltxmacro}
\usepackage{marginnote}
\LetLtxMacro{\oldtodo}{\todo}

\newif\ifAnon
\Anonfalse   

\definecolor{codebg}{RGB}{250,250,250}
\definecolor{codeframe}{RGB}{210,210,210}
\definecolor{kw}{RGB}{0,0,160}
\definecolor{cm}{RGB}{50,120,50}
\definecolor{st}{RGB}{160,20,20}
\definecolor{num}{RGB}{120,120,120}
\usepackage{tcolorbox}

\definecolor{step1}{HTML}{FF8000} 
\definecolor{step2}{HTML}{001DBC} 
\definecolor{step3}{HTML}{FF3333} 

\definecolor{ByteLoadColor}{HTML}{B85450}
\definecolor{MemBarrierColor}{HTML}{006600}
\definecolor{Delaycolor}{HTML}{6130A7} 

\lstdefinestyle{cstyle}{
  backgroundcolor  = \color{codebg},
  frame            = single,
  framerule        = 0.4pt,
  rulecolor        = \color{codeframe},
  basicstyle       = \ttfamily\scriptsize,
  keywordstyle     = \color{kw}\bfseries,
  commentstyle     = \color{cm}\itshape,
  stringstyle      = \color{st},
  numbers          = left,
  numberstyle      = \tiny\color{num},
  stepnumber       = 1,
  numbersep        = 5pt,
  showstringspaces = false,
  breaklines       = true,
  breakatwhitespace= true,
  captionpos       = b,
  tabsize          = 2,
  xleftmargin      = 1.5em,
  xrightmargin     = 0.3em,
}

\lstdefinestyle{asmstyle}{
  backgroundcolor  = \color{codebg},
  frame            = single,
  framerule        = 0.4pt,
  rulecolor        = \color{codeframe},
  basicstyle       = \ttfamily\scriptsize,
  commentstyle     = \color{cm}\itshape,
  numbers          = left,
  numberstyle      = \tiny\color{num},
  stepnumber       = 1,
  numbersep        = 5pt,
  breaklines       = true,
  captionpos       = b,
  tabsize          = 4,
  xleftmargin      = 1.5em,
  xrightmargin     = 0.3em,
}

\newtheorem{theorem}{Theorem}

\newcommand{\sysname}{\textsc{HOMARD}\xspace}

\newcommand{\tRCD}{\ensuremath{t_{\mathrm{RCD}}}\xspace}
\newcommand{\tRAS}{\ensuremath{t_{\mathrm{RAS}}}\xspace}
\newcommand{\tRP}{\ensuremath{t_{\mathrm{RP}}}\xspace}

\newcommand{\prettyparagraph}[1]{\textit{#1}}

\usepackage[hidelinks]{hyperref}
\usepackage[capitalize]{cleveref}
\hypersetup{pdftitle={HOMARD: Hammering Off-chip Memory via Aggregate power Rail Disclosure}, pdfsubject={Author preprint}}

\begin{document}

\title{HOMARD: Hammering Off-chip Memory via Aggregate power Rail Disclosure}

\ifAnon
  \author{Anonymous Author(s)}
\else
  \author{%
    \IEEEauthorblockN{%
      Eliott~Qu\'{e}r\'{e}\IEEEauthorrefmark{1},\;
          Li-Chung~Chiang\IEEEauthorrefmark{2},\;
      Carina~Fiedler\IEEEauthorrefmark{2},\;
      Thomas~Rokicki\IEEEauthorrefmark{1},\;
      Maria~M\'{e}ndez~Real\IEEEauthorrefmark{3}\\[2pt]
      Alessandro~Palumbo\IEEEauthorrefmark{1},\;
      Lilian~Bossuet\IEEEauthorrefmark{4},\;
      Rub\'{e}n~Salvador\IEEEauthorrefmark{1},\;
      Daniel~Gruss\IEEEauthorrefmark{2}%
    }
    \IEEEauthorblockA{\IEEEauthorrefmark{1}%
      CentraleSupélec, Univ.\ Rennes, Inria, CNRS, IRISA\\
      \{eliott.quere, thomas.rokicki, alessandro.palumbo, ruben.salvador\}@irisa.fr}
    \IEEEauthorblockA{\IEEEauthorrefmark{2}%
      Graz University of Technology\\
      \{li-chung.chiang, carina.fiedler, daniel.gruss\}@tugraz.at}
    \IEEEauthorblockA{\IEEEauthorrefmark{3}%
      Univ.\ Bretagne-Sud, Lab-STICC, UMR~6285\\
      maria.mendez-real@univ-ubs.fr}
    \IEEEauthorblockA{\IEEEauthorrefmark{4}%
      Univ.\ Lyon, UJM-Saint-\'{E}tienne, CNRS, Lab.\ Hubert Curien, UMR~5516\\
      lilian.bossuet@univ-st-etienne.fr}
  }
\fi

\IEEEaftertitletext{%
  \begin{center}
    \small Author preprint. Prepared for the 2027 IEEE Symposium on Security and Privacy (S\&P).
  \end{center}
  \vspace{0.5em}
}
\maketitle
\pagestyle{plain}
\thispagestyle{plain}

\begin{abstract}
For over a decade, software-based microarchitectural attacks have mainly exploited timing differences to infer information from DRAM.
However, recent timing-based approaches failed to characterize some precise microarchitectural behaviors, either due to defense or optimization mechanisms.

In this work, we show that unprivileged millisecond- and milliampere-resolution power sensors on modern Arm SoCs
are sufficient to recover DRAM mapping functions and reveal the data dependence of memory accesses without any timers.
We present \sysname, a new methodology that reads DRAM activity from this benign telemetry, overcoming the mentioned timing limitations.
Building on this, we introduce two new attack primitives.
\sysname-RE is a hierarchy-aware linear-algebraic method that reverse-engineers the DRAM addressing functions from power alone.
\sysname-DATA recovers the Hamming weight of a fetched cache line through joint-power rail templating, and the secret byte values that the CPU moves across the DRAM bus through correlation power analysis on the same rails.
Across an AMD ZCU102, a Raspberry Pi~5, and a multi-chip-based LPDDR4 NVIDIA Jetson~TX2, \sysname-RE succeeds where row-buffer timing methods fail.
On the Pi~5, \sysname-DATA classifies $\mathbf{9}$ Hamming-weight levels at $\mathbf{81.5\,\textbf{\%}}$ accuracy and recovers secret bytes in minutes by correlating the rail current with the data-dependent bit toggling between neighbors in a row.
Both primitives work with an architectural load or a mispredicted Spectre-RSB path on which no value is architecturally read.
Together, these results establish unprivileged power telemetry as an independent side channel for profiled attacks on the microarchitecture, particularly powerful when timing falls silent.
\end{abstract}

\section{Introduction}
\label{sec:intro}

Side-channel attacks have been studied for three decades, exploiting various signal sources and side effects such as timing, power, emissions of electromagnetic radiation, heat, and sound.
Many of these attacks require physical access whereas others, mostly timing attacks, enable software-based attacks, e.g., on microarchitectural elements as well.
Since software typically cannot make fine-grained observations on power consumption or physical emissions, most software-based microarchitectural attacks rely on timing differences.
However, timing attacks require high-resolution timers, which may be inaccessible~\cite{lipp2016armageddon}, or unable to capture fine-grained differences in the microarchitecture, such as the row-buffer conflict on some multi-chip LPDDR4 platforms~\cite{plin2026knockknock}.
Furthermore, timing attacks can only reveal differences that have different execution times in the microarchitecture, e.g., depending on the microarchitectural state or secret inputs to a computation.

Only recently, software-based power analysis has emerged, allowing to observe differences in the microarchitecture even when no timing difference exists, e.g., different values in a register.
This effect has been exploited on x86 processors to attack cryptography~\cite{lipp2020platypus}, or directly read data-dependent activity in the shared cache hierarchy~\cite{kogler2023collidepower} on x86.
These attacks exploit dedicated interfaces, e.g., Intel RAPL, and mechanisms, e.g., DVFS, to make these differences in the power consumption visible to a software-based attacker.
On Arm, userspace rail sensors have been used to fingerprint deep neural networks and recover cryptographic keys~\cite{zhang2025amperebleed,chawla2024uncovering}.
Similarly, custom circuits deployed in the programmable logic in an SoC-FPGA behave as delay sensors, providing indirect measurements of instantaneous power activity. SnoopyPower~\cite{quere2026snoopypower} exploited this behavior to attack the cache and the coherence states of the Arm complex in an AMD Zynq SoC.

In contrast to these attacks on the microarchitecture of the processor, software-based attacks on the DRAM still primarily rely on timing.
For instance, a long line of research exploits DRAM timings to infer physical-to-DRAM address mappings~\cite{pessl2016drama,Wang2020Dramdig,heckel2025verifying}.
However, DRAM activity is a dominant contributor to system-level power consumption: around 46\% on consumer platforms, and more than 60\% under memory-intensive workloads such as deep learning training~\cite{steiner2025drampower}.
Consequently, physical power side-channel attacks are known to be able to exploit power consumption variations in DRAM, which can be accumulated to leak cryptographic keys~\cite{mangard2007power}.
Therefore, it is surprising that there is only limited knowledge about the extent to which DRAM power consumption can expose system activities to software-based attackers, which have much more constrained access to power consumption data.
Furthermore, vendors have taken steps against prior software-based power attacks~\cite{lipp2020platypus}, by deploying mitigations in more recent processors.

In this paper, we present \sysname, an unprivileged software-based power side channel that leverages low-resolution sensors on modern Arm-based platforms.
We develop a methodology for reverse-engineering DRAM addressing functions from the hierarchy and its levels of contention.
We show that power consumption can overcome limitations of timing methods and resolve the hierarchy at a finer granularity.
In particular, our method allows us to recover the mapping from systems where the state of the art failed with timing-based methods~\cite{plin2026knockknock}.
We turn this into a generic, channel-independent recovery method that, for the first time, exploits the hierarchy's nesting to reduce the number of address pairs required for reverse engineering, and into which any channel could plug.
Power channels make the method especially practical, since they expose the hierarchy levels directly.
On an Arm SoC with multiple banks, sub-channels, and two separate chips, our method recovers the full mapping and reduces the number of address pairs needed by 2.15$\times$ over the state of the art in DRAM reverse engineering.

We demonstrate, for the first time, that DRAM power consumption, read via a software interface exposed to unprivileged entities, \emph{can leak the value of a memory block loaded from DRAM}, with 63 out of 64 bytes recovered. By combining this leakage with speculative execution, we can even extract the value of an arbitrary memory block from a victim's address space with the same accuracy.

We first identify DRAM power rails that expose power consumption to unprivileged entities on several Arm-based SoCs.
We then characterize the correlation between power consumption and the content from a memory block (a 64-byte-aligned memory region) accessed from the DRAM.
We show that power consumption correlates with the number of bits that flip between two memory blocks accessed back-to-back, i.e., their Hamming distance.
Building on this observation, we present a novel attack technique that induces concurrent accesses to a secret memory block and an attacker-controlled memory block. By varying the attacker-controlled data and measuring the resulting power consumption, the attacker can apply correlation power analysis (CPA) to recover the exact value of the secret memory block.
Our attack recovers 63 bytes of a 64-byte secret memory block correctly in 19 hours.
Finally, by combining this technique with speculative execution, we show that an attacker can induce speculative accesses to an arbitrary memory block in the victim's address space and achieve the same accuracy within 18 hours.

\subsection*{Contributions}
We establish non-privileged, aggregate \textbf{power} telemetry \textbf{as a general-purpose microarchitectural side channel on Arm SoCs}. Through \sysname, we contribute:
\begin{enumerate}[leftmargin=1.5em, noitemsep, topsep=3pt]
  \item \textbf{Timer-free DRAM reverse-engineering.} \sysname exploits a power side-channel to recover the DRAM hierarchy and the functions mapping physical addresses to it, using userspace telemetry alone.
  \sysname enables recovering, for the first time, the full DRAM hierarchy on Arm systems; this yields more information than state-of-the-art tools and allows to enhance DRAM attacks.
  Our methodology allows us to retrieve the DRAM mapping of multi-chip LPDDR4 systems, where state-of-the-art timing-based approaches failed.

  \item \textbf{Hierarchy-aware reverse-engineering.} We introduce a new methodology that extends the binary oracle (i.e., row-buffer conflict or miss) of prior work to a \emph{multi-level} oracle, yielding more information on the relation between the addresses.
  By exploiting the nested hierarchy, it replaces a single exponentially rare collision with a sequence of cheaper conditional events, reducing the number of address pairs the probe requires.

  \item \textbf{DRAM Data Leakage.} We demonstrate the first software-only power analysis attack that can leak DRAM data. \sysname can not only derive the Hamming weight of a memory block but also leak its content using CPA techniques.

  \item \textbf{Power-based Spectre-RSB.} We mount an end-to-end Spectre-RSB attack using \sysname as the covert channel to leak information about arbitrary data from the victim's address space. Unlike traditional cache-based Spectre attacks, our attack does not require a secret-dependent memory access pattern, broadening the range of vulnerable applications.
\end{enumerate}

\prettyparagraph{Responsible Disclosure and Open Access.}
We disclosed our findings to the national CERT on June 09, 2026, and to Raspberry Pi Ltd, the company behind the Raspberry Pi platform, on June 10. We are in the process of disclosing our findings to other affected vendors with the assistance of the national CERT. All code and data are available at
\href{https://doi.org/10.5281/zenodo.23078196}{Zenodo (DOI: 10.5281/zenodo.23078196)},
mirrored at
\href{https://github.com/Koyiott/homard}{github.com/Koyiott/homard}.


\section{Background and Related Work}
\label{sec:background}

\subsection{DRAM Fundamentals}
\label{sec:background:dram-fundamentals}
\begin{figure}[t]
  \centering
  \includegraphics[width=1\linewidth]{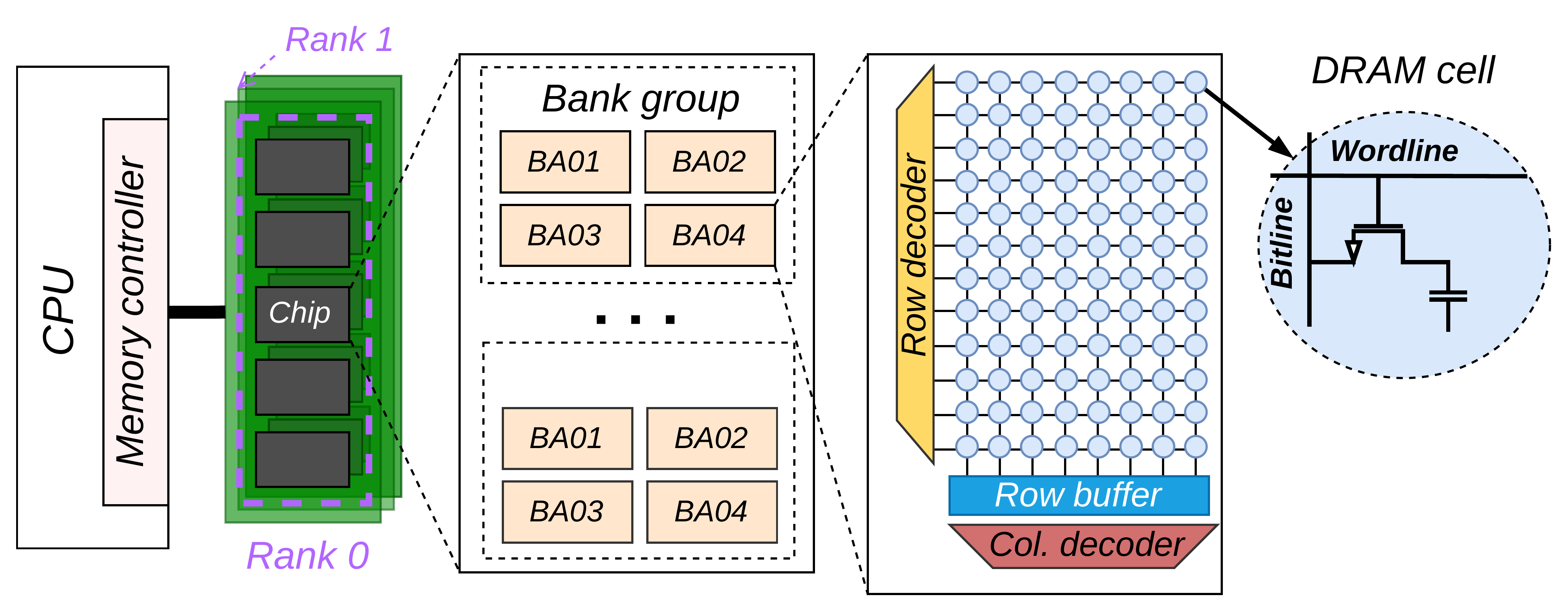}
  \caption{DRAM-based memory system organization.}
  \label{fig:memory_hierarchy}
\end{figure}

To increase bandwidth and scheduling parallelism, DRAM is organized as a hierarchy of channels, ranks, chips, banks, rows, and 
more recently bank groups~\cite{steiner2025drampower,kim2012case}.
Figure~\ref{fig:memory_hierarchy} summarizes this organization:
a memory channel connects the controller to one or more ranks, each containing multiple chips that operate in lockstep to serve a cache-line access.
Within each chip the banks process commands with a degree of independence while still sharing channel-level resources such as the command/address and data buses~\cite{steiner2025drampower}.

At the array level, each cell stores one bit as charge on a capacitor.
The controller first issues an \texttt{ACT} command to open a row in a bank; the activated row is sensed and latched in the row buffer, after which \texttt{RD} and \texttt{WR} commands select columns and transfer data in bursts over the memory interface~\cite{keeth2007dram}.
Because only one row can be active per bank at a time, a request to a different row in the same bank requires a \texttt{PRE} command to close the current row before another \texttt{ACT} can be issued, resulting in a \emph{row-buffer conflict}.
Cell charge leaks over time, so the controller must also issue refresh commands periodically; depending on the standard, refresh may target all banks or only a subset, temporarily blocking access to the refreshed banks.%

Modern DRAM uses a bidirectional data bus with data strobes, a separate command/address bus, and high-speed signaling with receiver termination.
Current-generation interfaces operate at multi-Gb/s per pin and must account for termination, channel loss, and package topology~\cite{hollis2019recent,jouppi2012cacti}.


%
For power and energy accounting, DRAM standards specify IDD current conditions for fixed operating scenarios.
These classes formalize a command-level view of DRAM activity, tying each hierarchy interaction to a distinct mix of activation, precharge, read, write, refresh, standby, and power-down states.
Measurement conditions differ across DDR, LPDDR, GDDR, and HBM standards, and accurate power accounting must capture active-bank state, command counts, and interface behavior rather than rely on a single uniform per-access cost~\cite{steiner2025drampower,jung2016new,ghose2018your}.

\subsection{Remote Power Analysis in SoC}
\label{sec:background:power}
Power analysis exploits the data dependence of CMOS power consumption: each logic transition charges or discharges capacitance, and switching activity dominates instantaneous current~\cite{rabaey2003digital,mangard2007power}.
The majority of remote power side-channel work relies on non-profiled attacks, recovering secrets from instantaneous or correlated current traces via simple power analysis (SPA), differential power analysis (DPA), and correlation power analysis (CPA)~\cite{kocher1999differential,brier2004correlation}.
In contrast, profiled (template-based) analysis~\cite{standaert2009unified}
has only recently been extended beyond cryptographic targets to fingerprint microarchitectural states themselves~\cite{quere2026snoopypower}, and the body of work that does so on commodity SoCs remains comparatively small.

This shift has been enabled by software-accessible power telemetry, which turns template-based observation into a purely software primitive on modern SoCs~\cite{le2025not}.
On x86, Intel's RAPL counters enabled practical key recovery against SGX enclaves in PLATYPUS~\cite{lipp2020platypus}.
Collide+Power moved the methodology into the microarchitectural domain by showing that data-dependent activity in the shared memory hierarchy leaves a signature in coarse software-readable energy interfaces, with no cryptographic target required on the victim side~\cite{kogler2023collidepower}.
On Arm, userspace rail sensors fingerprint DNN models and leak RSA-1024 keys~\cite{zhang2025amperebleed}, and the Apple M1/M2 SMC's on-chip meters yield AES-128 keys via unprofiled CPA~\cite{chawla2024uncovering}.
Microarchitectural power templating on Arm exists but requires an instantiated Time-To-Digital Converter (TDC) rather than native SoC telemetry~\cite{quere2026snoopypower}, and, like sensor-based attacks, targets only the compute side.
DRAM contention and its imprint on the dedicated memory-subsystem rails, already exposed to userspace on production Arm SoCs, remain largely unexplored.

\subsection{Software-Based DRAM Address Reverse Engineering}
\label{sec:background:dram}
The physical-to-DRAM mapping is a prerequisite for reliable side-channel~\cite{kwong_rambleed_2020} and fault-injection analyses~\cite{Frigo2020TRRespass}, as well as for constructing and mitigating attacks such as Rowhammer~\cite{ZebRAM}.
An extensive body of literature aims at reversing it.
DRAMA~\cite{pessl2016drama} first showed how to recover undocumented physical-to-DRAM mapping purely from software: it uses the row-buffer-conflict timing channel to group addresses by bank, then brute-forces XOR masks that reproduce each group. This search is exponential in the number of address bits. Later work~\cite{Xiao2016, Wang2020Dramdig, Frigo2020TRRespass, Heckel2023Reverse-Engineering, Marazzi2024risc, Jattke2024Zenhammer, Chen2025RowhammerPrefetch, wi2025sudoku, plin2026knockknock} builds on the same timing channel. 
The approach has been ported to other platforms: AMDRE~\cite{Heckel2023Reverse-Engineering} and  DARE~\cite{Jattke2024Zenhammer} target AMD, while RISC-H~\cite{Marazzi2024risc} targets RISC-V.
Several works reduce the runtime by testing strategically selected address pairs~\cite{Xiao2016, Chen2025RowhammerPrefetch} or speeding up the exponential search using DRAM geometry and microarchitectural priors~\cite{Wang2020Dramdig}. 
Knock-Knock~\cite{plin2026knockknock} avoids brute force and recovers the masks as a null space in polynomial time.
Sudoku~\cite{wi2025sudoku} adds two more timing channels to label each mask by hierarchy level.
A separate approach drops timing altogether and reads the channel, rank, and bank hierarchy from Intel performance counters~\cite{helm2020reliable}.

All of these methods either rely on a latency gap that only a fine-grained timer can measure or depend on vendor-specific performance counters.
The weakness of timing-based approaches is that the latency gap becomes hard to exploit on modern hardware:
the gap is small and noisy on multi-chip LPDDR4 parts~\cite{plin2026knockknock}; it is coarsened and jittery in sandboxes and browsers~\cite{deRidder2021}; and it disappears on controllers that equalize access latencies. 

\subsection{Row-Conflict Timing Side Channel Limitations}
\label{sec:channel:timing-limits}
A DRAM access decomposes into a sequence of array operations whose latencies are exposed through standard timing parameters such as \tRCD, \tRAS, and \tRP~\cite{chang2016understanding}.
\tRCD bounds the delay between an \texttt{ACT} command and the first safe access to the activated row, \tRAS specifies the time required to restore the row, and \tRP captures the precharge delay before another row in the same bank can be activated.
As a result, row-buffer hits, same-bank row conflicts, and accesses that proceed in parallel across banks or channels fall into distinct latency classes that an adversary can exploit to recover the underlying address mapping.

This latency signal is informative, yet it is far from a clean function of the address mapping alone.
Standard timings are derived from worst-case margins, whereas real devices exhibit substantial variation across cells and spatial regions~\cite{chang2016understanding}.
Physical layout amplifies this variation: cells at different distances from sense amplifiers, wordline drivers, and other peripheral structures experience different RC delays, so some internal regions can be accessed faster than others~\cite{lee2017design}.
In addition, recently accessed rows may remain highly charged after precharge and can tolerate reduced \tRCD{} and \tRAS{}, showing that timing also depends on transient row history, leakage, and refresh state~\cite{hassan2016chargecache}.

The memory controller reorders pending references to improve locality and bus utilization~\cite{rixner2000memory}.
A measurement, therefore, mixes the probed accesses with queuing, refresh, prefetching, and OS noise.
Extracting the row-conflict signal from this mixture requires
high-resolution timers and careful calibration, and that dependence is
the limitation: timers can be coarsened or removed, and the few methods
that instead use performance counters~\cite{helm2020reliable} can be restricted just as easily, so a simple defense disables the existing channels. Even when the timer is available, on Arm devices, the channel remains binary, reporting only whether two addresses collide at the deepest level and discarding higher levels of contention in the hierarchy.
This discarded information is increasingly leveraged: early Rowhammer attacks primarily need knowledge of collisions at the deepest bank level, whereas recent approaches~\cite{baek2025marionette} target specific DRAM components.
Hence, higher-level co-location may enable future attacks and improve their performance.

\section{Threat Model}
\label{sec:threat:attacker}
We consider an adversary with two complementary goals on a target platform.
In the first goal, the attacker attempts to recover the DRAM memory hierarchy and its associated address mapping function, expressed as Boolean functions that, given a physical address, determine its chip, sub-channel, and bank.

In the second goal, the attacker aims to exfiltrate data-dependent state from memory blocks accessed from the DRAM. Such accesses include not only architectural accesses but also the blocks accessed along transient execution paths. The transient access is often triggered in Spectre attacks. Both goals rely on the same online primitive introduced in \autoref{sec:channel:wfp}; they differ only in the stimulus and the measurement objectives (physical location or data). All prior software-level recovery techniques against either goal assume the adversary can measure latency with sub-microsecond resolution; we remove this assumption entirely.

\prettyparagraph{Platform and security domains.}
\sysname assumes a commodity multi-user OS and unprivileged code execution: no root, no physical access. In particular, the attacker is an unprivileged party on the same SoC, i.e., any third-party application or another tenant. The victim can be any other process regardless of its running privileges, such as a kernel or cryptographic processes.
Our threat model considers platforms that expose the considered telemetry sensors to unprivileged user code. These are the types of platforms typically deployed in embedded domains, such as automotive, industrial, or other edge computing scenarios. Some examples include embedded GPU-based platforms like the Jetson family, multi-processor SoCs with FPGA like AMD Ultrascale+ MPSoC (e.g., as in the ZCU102 board), or Raspberry~Pi~5 SoCs. These platforms, which include application-grade processors, run a typical commodity multi-user OS such as Linux.

\subsection{Online Capabilities}
\label{sec:threat:capabilities}
The adversary can run ordinary user-space processes on the target platform,
with no root permissions, no kernel modules, and no \texttt{CAP\_SYS\_*}
capabilities. The attacker requires three primitives:
(i) allocate memory and issue loads to chosen virtual addresses;
(ii) evict cache lines through the userspace flush instruction
(\texttt{DC~CIVAC} on Arm); and (iii) read a \emph{non-privileged power
telemetry interface} exposing a scalar proxy for aggregate current or
power over a domain that includes the DRAM subsystem, sampled at low
timing resolution. The attacker does not require a high-resolution timer, privileged platform control (governor, thermal state, DVFS), nor physical access or external instrumentation.

We do not restrict ourselves to a particular type of power sensor.
We use the accessible power telemetry that the platform exposes and adapt the primitive to it.
In modern SoCs, for example, the operating system exposes structurally equivalent interfaces through PMIC mailboxes, Linux \texttt{hwmon}, and IIO drivers, with analogous interfaces available on x86 (RAPL) and server platforms (BMC (Baseboard Management Controller) rails).
Access policy varies: on some platforms, the relevant device node is world-readable by default; on others, it is reachable through membership in a standard, non-administrative group (e.g.,\ \texttt{video}, \texttt{gpio}) that ordinary users routinely hold.
In every case, the requirement is weaker than \texttt{root}, any \texttt{CAP\_SYS\_*} capability, or the timer access prior work assumes~\cite{pessl2016drama,plin2026knockknock,wi2025sudoku}.


\subsection{Physical-Address Visibility and its Consequences}
\label{sec:threat:pa}
Recovering the mapping functions means observing how differences in physical address sort probe pairs into the DRAM hierarchy. How much of the physical address userspace can see, therefore, bounds what the attacker can recover on its own, giving rise to two regimes.

\prettyparagraph{Self-sufficient userspace.}
If the OS hands out large, aligned, contiguous pages, the virtual and physical addresses agree on every bit the hierarchy uses ($\mathrm{VA}[i]=\mathrm{PA}[i]$), and the full mapping follows from virtual addresses alone.

\prettyparagraph{Hierarchy observation.}
If the mapping reaches past the large-page range, userspace can no longer see the high physical bits needed to fully invert it. It still recovers the hierarchy's structure from virtual-address pairs: how many contention levels exist and how often pairs fall into each. What it cannot do is \emph{name} those levels in physical-address terms. This requires a single offline profiling step via a privileged translation interface (\texttt{/proc/self/pagemap}), whose result is fixed by the DDR-controller settings and reused indefinitely. Prior software methods pay the same offline cost on the same interface~\cite{pessl2016drama,plin2026knockknock,wi2025sudoku}, so we claim no lower privilege; what we remove is the timer, reading the hierarchy from power instead~\cite{pessl2016drama,plin2026knockknock}.
Most importantly, the information exposed by the power channel is richer, resolving hierarchy levels from contention, where the timing channel fails. 

\section{DRAM Contention Power Side Channel}
\label{sec:channel}
This section introduces the channel at the core of \sysname.
A stream of memory operations drives the subsystem through state transitions whose energy costs differ, and every rail that supplies the memory subsystem integrates these costs into a current the attacker can read.
Fully unprivileged and timerless, \sysname separates execution paths by the current they draw.
It is sensitive to the structural state, e.g., the DRAM hierarchy, as well as to the data.
\autoref{sec:channel:physics} develops the physical mechanism that links memory activity to rail current, and
\autoref{sec:channel:wfp} formalizes the measurement primitive.

\subsection{Physical Mechanism}
\label{sec:channel:physics}
A DRAM access is a sequence of electrical events. Each command, \texttt{ACT}, \texttt{PRE}, \texttt{RD}, and
\texttt{WR}, draws its own current according to the work it does inside
the chip. Total power depends on the command stream, the active bank
state, and the data moved, not on a uniform per-access
cost~\cite{vogelsang2010understanding, ghose2018your}.

Modern DRAM splits its supplies into several voltage domains, such as the
core array, the boosted wordline, and the DQ I/O, each exposed on a
separate rail that a board-level sensor reads
individually~\cite{steiner2025drampower, micron_lpddr4x_2023}. The current
on these rails tracks the command mix, from tens of milliamps at standby
to hundreds during a read burst~\cite{ghose2018your}.

Two axes modulate this current. The first is \emph{command density}, the
mix and rate of commands the access pattern imposes. A same-bank conflict
adds a \texttt{PRE} and \texttt{ACT} per access and serializes the
schedule; bank- or channel-parallel patterns sustain a denser stream.
Command density depends on where the accessed addresses meet in the DRAM
hierarchy, and \sysname-RE reads it (\autoref{sec:dramre}).

The second is \emph{data-dependent current}, set by the bits the access
moves~\cite{ghose2018your}. The read current carries two linear terms: it
rises with the Hamming weight of the line, and with the number of bits
that toggle between two consecutive transfers, the Hamming distance.
\sysname-DATA reads these terms (\autoref{sec:dramhw}).

The available power sensor never exposes the rail current at this
timescale. Internally, it samples at a high rate~\cite{ti_ina226_2024}, but it does not report those samples: each user-visible value is an integration of many internal
samples~\cite{linux_ina2xx_2024}. Modeling one conversion as an integrate-and-dump, the reported
sample is the mean current over a window $T$,
\begin{equation}
  y[k] \;\approx\; \frac{1}{T}\int_{kT}^{(k+1)T} i(t)\,dt,
  \label{eq:boxcar}
\end{equation}
which already averages away anything faster than the window. The sensor
then averages $N$ such conversions before updating its registers, so the
user-visible interval $T_{\mathrm{upd}}\approx N\,T$ stretches to
milliseconds, and the exported rate drops to tens of hertz~\cite{linux_ina2xx_2024}.
Userspace therefore reads a heavily low-pass-filtered mean of the rail current, far slower than any single command could be, and never the instantaneous power.

This two-stage path, fast internal sampling followed by integration and
averaging, is common to the shunt monitors and PMICs we use. The Pi~5 PMIC~\autoref{fig:setup:pi5} lacks a public datasheet, but our experiments suggest it follows the same model. The integration that hides individual events is exactly what we exploit: repeating a stimulus accumulates its contribution inside the
window of \autoref{eq:boxcar}, so the same averaging that suppresses a
one-shot access amplifies a sustained one. \autoref{sec:channel:wfp}
builds the primitive on this.

A single command or bit transition lasts nanoseconds and is erased by this
window, so neither axis appears in one access. The readable quantity is
the windowed mean of \autoref{eq:boxcar}. The two axes are orthogonal, so
we isolate one by keeping the other constant: with the data content
identical across accesses, only command density varies and the current
exposes the hierarchy (\autoref{sec:dramre}); with the access pattern
identical, only the data varies and the current exposes the content of a
memory block (\autoref{sec:dramhw}). One channel thus carries both
attacks, and \autoref{sec:channel:wfp} builds the primitive that fills
this window to read it.

\subsection{The \sysname Primitive}
\label{sec:channel:wfp}
\sysname rests on a single primitive: a loop that repeats a stimulus
$\mathcal{H}$ for $N$ iterations and reads the current the memory
subsystem draws while it runs. The power sensors exposed to the userspace are coarse and slow. One flush-and-load swings the rail by microamps over nanoseconds, far below a telemetry interface that samples at tens of hertz. Repeating $\mathcal{H}$ integrates its per-iteration contribution over the sensor window until it rises above the noise floor (\autoref{sec:channel:physics}). Repetition is the amplifier that turns a low-resolution power sensor into a reliable side-channel against DRAM microarchitecture.

The iteration count $N$ is the only knob. We raise it until the per-probe deviation falls below one sensor LSB, so the same loop runs unchanged across platforms that differ in sensor resolution, sampling rate, and noise floor; the attacker sets $N$ from the SNR that each platform and each target demands (\autoref{sec:dramre:setup}). The loop returns the rail current averaged over the $N$-iteration burst (\autoref{alg:primitive}); each attack then maps that scalar to its label.

\begin{algorithm}[h!]
\footnotesize
\caption{The \sysname primitive.}
\label{alg:primitive}
\begin{algorithmic}[1]
\Require Stimulus $\mathcal{H}$; iteration count $N$
\Ensure Mean rail current $\mu$ over the burst
\Function{Homard}{$\mathcal{H},\ N$}
  \State \textcolor{step1}{\Call{Start\_Sampler}{ }}
  \State \textcolor{step2}{\Call{Stimulate}{$\mathcal{H},\ N$}}
  \State \Return \textcolor{step3}{$\mu \gets$ \Call{Mean\_Rail}{ }}
\EndFunction
\end{algorithmic}
\end{algorithm}

The hammering stimulus $\mathcal{H}$ is generic for each attack. We instantiate it twice, once to recover \emph{where} memory lives and once to recover \emph{what} the controller moves.

A \emph{structural} stimulus $\mathcal{H}_{\mathrm{pair}}(A,B)$ is an address pair. Each iteration flushes and reloads both addresses, and the command stream the controller issues (\texttt{ACT}, \texttt{PRE}, \texttt{RD}) depends on where $A$ and $B$ meet in the DRAM hierarchy. The returned scalar therefore recovers the hierarchy relation, the bank, sub-channel, and chip selectors (\autoref{sec:dramre}).

A \emph{data-dependent} stimulus $\mathcal{H}_{\mathrm{D}}$ is a memory access whose content fixes the bus-toggle profile of the transfer. The returned scalar then recovers the content moved over the bus, the Hamming weight of a line or its exact bytes (\autoref{sec:dramhw}).

The two attacks share one loop and one amplifier; only the stimulus and $N$ change, and a classifier, template, or correlation maps the returned scalar to its class. Used structurally, the primitive resembles the classical row-conflict channel~\cite{pessl2016drama}, but reads power rather than latency and separates each level of the hierarchy instead of a single same-bank bit. Used on data, it has no counterpart in prior software-only channels on the platforms we study.

\section{HOMARD-RE: DRAM Hierarchy Recovery}
\label{sec:dramre}
\sysname-RE instantiates the primitive of \autoref{sec:channel:wfp} with the structural stimulus $\mathcal{H}_{\mathrm{pair}}(A,B)$, recovering the DRAM address-mapping hierarchy on Arm without timers or performance counters. Each iteration issues memory loads to the address pair, and the controller answers with a command sequence that depends on where $A$ and $B$ meet in the hierarchy: a row-buffer conflict on a shared bank, a shared sub-channel or chip, or no shared structure at all. Repeating the stimulus integrates this command-density difference into a stable rail current, which a lightweight classifier maps to a hierarchy level. Because the current reflects every level at which the two addresses agree, not only whether they collide at the deepest one, the oracle is multi-level rather than binary, exposing a structure that timing channels on the same controller cannot resolve~\cite{pessl2016drama}.

\subsection{Attack Design}
\label{sec:dramre:design}
We instantiate the \sysname primitive with a structural stimulus: an
ordered address pair $(A, B)$ whose relation in the DRAM hierarchy is
unknown to the attacker. Each iteration flushes the cache lines
containing $A$ and $B$ and issues memory loads to both. The
controller's response depends on whether $A$ and $B$ collide on a bank,
share a sub-channel or a chip, or share no structure at all: each case
produces a different mix of \texttt{ACT}, \texttt{PRE}, \texttt{RD}, and
\texttt{WR} commands per iteration. Repeating the stimulus over $N$
iterations integrates this command-density difference into a stable mean
rail current $\mu(A, B)$. We obtain $\mu$ by calling \textsc{Homard} with
the structural stimulus $\mathcal{H}_{\mathrm{pair}}(A, B)$, then a
lightweight classifier maps $\mu$ to a hierarchy label $L \in
\{\mathrm{conflict}, \mathrm{subch}, \mathrm{chip}, \mathrm{distinct}\}$;
we call this probe \textsc{Pair-Probe}.

The attacker runs \textsc{Pair-Probe} on a structured set of address pairs $\{(A_i, B_i)\}_{i=1}^{M}$ and accumulates the labelled XOR differences
\begin{equation*}
\mathcal{D} = \{(A_i \oplus B_i, L_i),\
  L_i = \textsc{Pair-Probe}(A_i, B_i, N)\},
\end{equation*}
which is the input to the recovery procedure of
\autoref{sec:dramre:recovery}.

Two properties of this instantiation matter for what follows. First, the
output is multi-level by construction: $\mu$ reflects every level at
which $A$ and $B$ may agree, not only the deepest one, so the oracle
returns one of four labels rather than a same-bank bit. Second, the
stimulus fixes the data pattern across all probes, isolating
command-density variation as the dominant source of signal.

\subsection{Hierarchy-Aware Reverse Engineering}
\label{sec:dramre:recovery}

\subsubsection{From Binary to Multi-Level Oracles}
\label{sec:dramre:recovery:oracle}
The response of \textsc{Pair-Probe} to an address pair $(A, B)$ is not a binary same-bank bit but a label
$L_i$ identifying the deepest level at which $A$ and $B$ agree in the DRAM hierarchy. This richer output is a structural property of the attack, not of the rail-current channel: any side channel whose
observable separates two addresses by the level of their agreement, rather than only by whether they collide at the deepest level, exposes the same labels. We develop the recovery procedure for such a \emph{multi-level oracle} abstractly in this section, and evaluate the specific case of \sysname's 4-level rail-current oracle.

Let $V = \mathbb{F}_2^n$ be the candidate physical-address space. A
hierarchy with $r$ levels is described by selector matrices
$H_1, \ldots, H_r$, where level $j$ contributes $q_j = K_j - K_{j-1}$
new linearly independent selector dimensions (with $K_j =
\mathrm{rk}(H_{\le j})$ and $K_0 = 0$). Two addresses $A$ and $B$ agree
through level $j$ exactly when $H_{\le j} D^\top = 0$, where $D = A
\oplus B$. A multi-level oracle on input $D_i$ returns the largest $j$
satisfying this condition; equivalently, $L_i = j$ encodes the $j-1$
nested kernel memberships
\[
D_i \in \ker(H_{\le 1}) \supset \ker(H_{\le 2}) \supset \cdots \supset \ker(H_{\le L_i}).
\]
A binary same-bank oracle is the degenerate case $L_i \in \{r, \neg r\}$
that collapses all intermediate equalities into a single bit.

\subsubsection{Sequential Recovery}
\label{sec:dramre:recovery:procedure}
The multi-level oracle admits a recovery that proceeds one hierarchy
level at a time, from the most probable equality to the least probable
one. At stage $j$, previously recovered selectors define the sampling
space $\ker(H_{<j})$. The attacker generates uniform differences in this
space, queries the oracle, and keeps the pairs satisfying
$L_i \ge j$. Stacking those differences into a matrix $D_j$ and solving
$D_j x^\top = 0$ yields, at full rank,
\[
\mathrm{nullspace}(D_j) = \mathrm{span}(H_{\le j}).
\]
The new selectors introduced at level $j$ are recovered as any
complement basis of $\mathrm{span}(H_{<j})$ inside this null space.

Two properties make this procedure correct and efficient. First, the
recovered span at stage $j$ is invariant under bit reuse across
hierarchy levels: $H_j$ may reuse address bits that already appear in
$H_{<j}$, and only the rank increase $q_j$ matters. Second, conditional
on exact recovery at stage $j-1$, the stage-$j$ samples are uniform on
$\ker(H_{<j})$, so the filtered useful samples are uniform on
$\ker(H_{\le j})$, a subspace of dimension $n - K_j$.

\subsubsection{Sample Bound}
\label{sec:dramre:recovery:bound}
The cost of recovery depends on how often the oracle observes the next
equality. Inside $\ker(H_{<j})$, the conditional probability of agreement
at level $j$ is
\[
\Pr[H_j D^\top = 0 \mid H_{<j} D^\top = 0]
  = \frac{|\ker(H_{\le j})|}{|\ker(H_{<j})|} = 2^{-q_j}.
\]
A binary oracle targeting only the deepest equality observes this event
with probability $\prod_{j=1}^{r} 2^{-q_j} = 2^{-K_r}$ in the full
address space, paying the product of all level probabilities as its
collision factor. A hierarchy-aware oracle replaces this single rare
event with $r$ conditional events, each sampled inside the kernel
recovered at the previous stage.

\begin{theorem}[$r$-level sample bound]
\label{thm:dramre:bound}
Let an oracle return the deepest agreement level $L_i \in \{1, \ldots,
r\}$ for each pair, with per-stage classification noise $\theta_j$ and
per-stage failure budgets $\varepsilon_j > 0$ summing to $\varepsilon$.
Hierarchy-aware recovery requires in expectation
\[
N_{\mathrm{seq}} \;\ge\;
  \sum_{j=1}^{r} \frac{2^{q_j}}{1 - \theta_j}
  \log_2\!\left(\frac{r\,(2^{n - K_j} - 1)}{\varepsilon}\right)
\]
measurements, against
\[
N_{\mathrm{bin}} \;\ge\;
  \frac{2^{K_r}}{1 - \theta}
  \log_2\!\left(\frac{2^{n - K_r} - 1}{\varepsilon}\right)
\]
for a binary same-bank oracle. The per-trace improvement is the ratio
\[
\sum_{j=1}^{r} 2^{q_j} \big/ 2^{K_r},
\]
which is strictly less than one whenever $r \ge 2$ and at least two
levels satisfy $q_j \ge 1$.
\end{theorem}

\prettyparagraph{Why hierarchy-awareness.}
The binary approach pays a single exponential factor $2^{K_r}$ in
collision rate; the hierarchy-aware approach pays the sum
$\sum_j 2^{q_j}$. The additive $\log_2 r$ overhead from the union bound
is asymptotically dominated by this multiplicative saving. The
improvement compounds with hierarchy depth: each additional level adds
$2^{q_j}$ to the sum but multiplies the product. We validate
the bound empirically on real hardware in \autoref{sec:dramre:recovery:bound}.

\prettyparagraph{Channel-agnosticism of the methodology.}
\autoref{sec:dramre:recovery:bound} makes no reference to the physical mechanism
that produces the labels $L_i$. The procedure of
\autoref{sec:dramre:recovery:procedure} applies to any oracle whose
classification noise satisfies $\theta_j < 1$ at every hierarchy level
the attacker wishes to recover; the per-stage cost degrades gracefully
as $\theta_j$ increases. In particular, the methodology recovers the
binary case as the special instance in which the oracle exposes only
the deepest level, reproducing the bound
of~\cite{plin2026knockknock}. The rail-current channel of \sysname
happens to expose four levels with $\theta_j \approx 0$; we evaluate
this specific instantiation in \autoref{sec:dramre:recovery-results}.

\subsection{Platforms and Setup}
\label{sec:dramre:setup}
We evaluate \sysname on three architecturally distinct platforms (the
AMD ZCU102, the Raspberry~Pi~5, and the NVIDIA Jetson~TX2), each
pairing a different controller architecture with a different commodity
power-telemetry interface. This diversity exercises \textsc{Pair-Probe}
under varying bank arithmetic, hierarchy depth, sensor LSB, and
sampling rate without changes to the primitive itself.

\begin{figure}[h]
  \centering
  \begin{subfigure}[t]{0.4\linewidth}
    \centering
    \includegraphics[width=\linewidth]{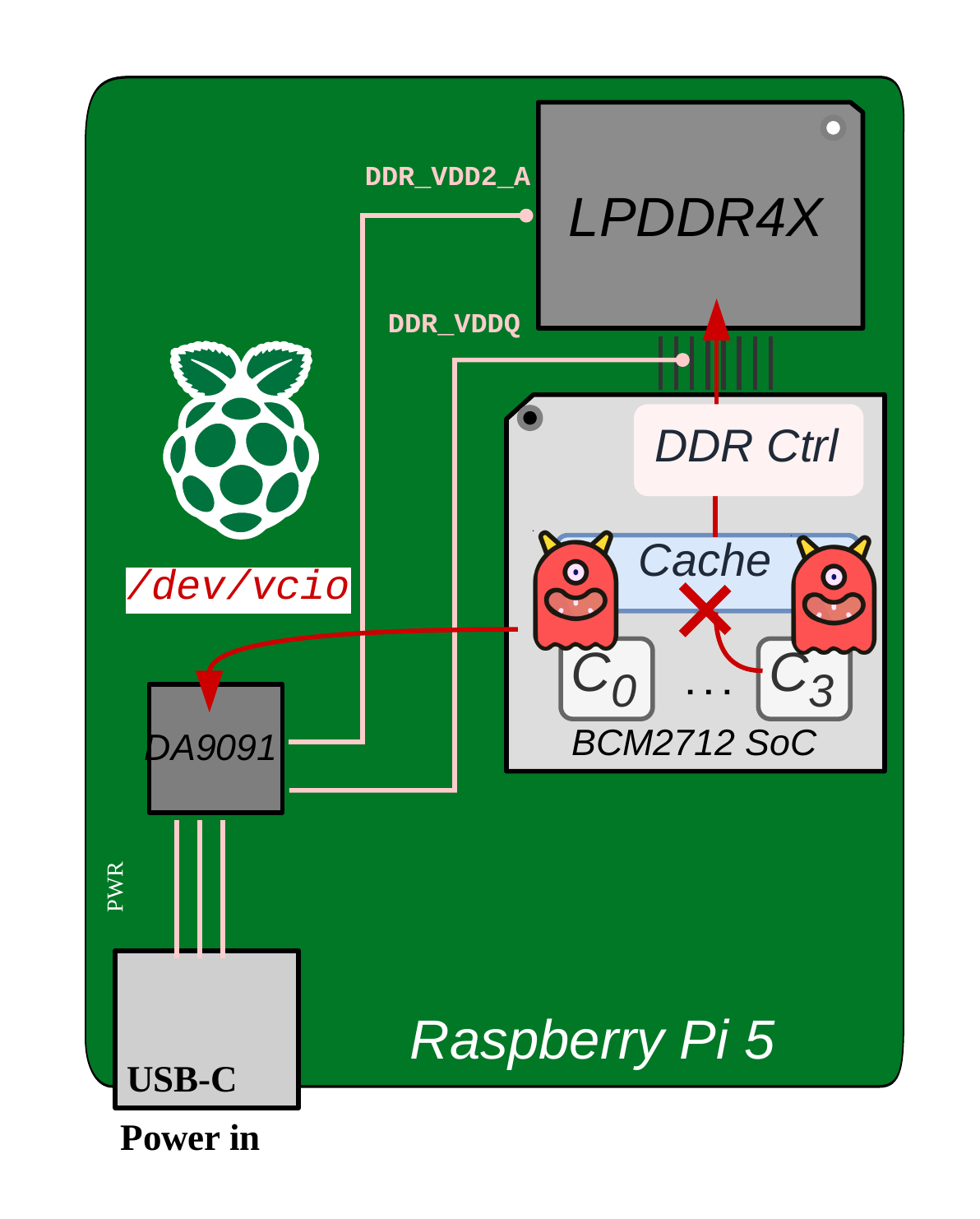}
    \caption{Raspberry~Pi~5.}
    \label{fig:setup:pi5}
  \end{subfigure}\hfill
  \begin{subfigure}[t]{0.55\linewidth}
    \centering
    \includegraphics[width=\linewidth]{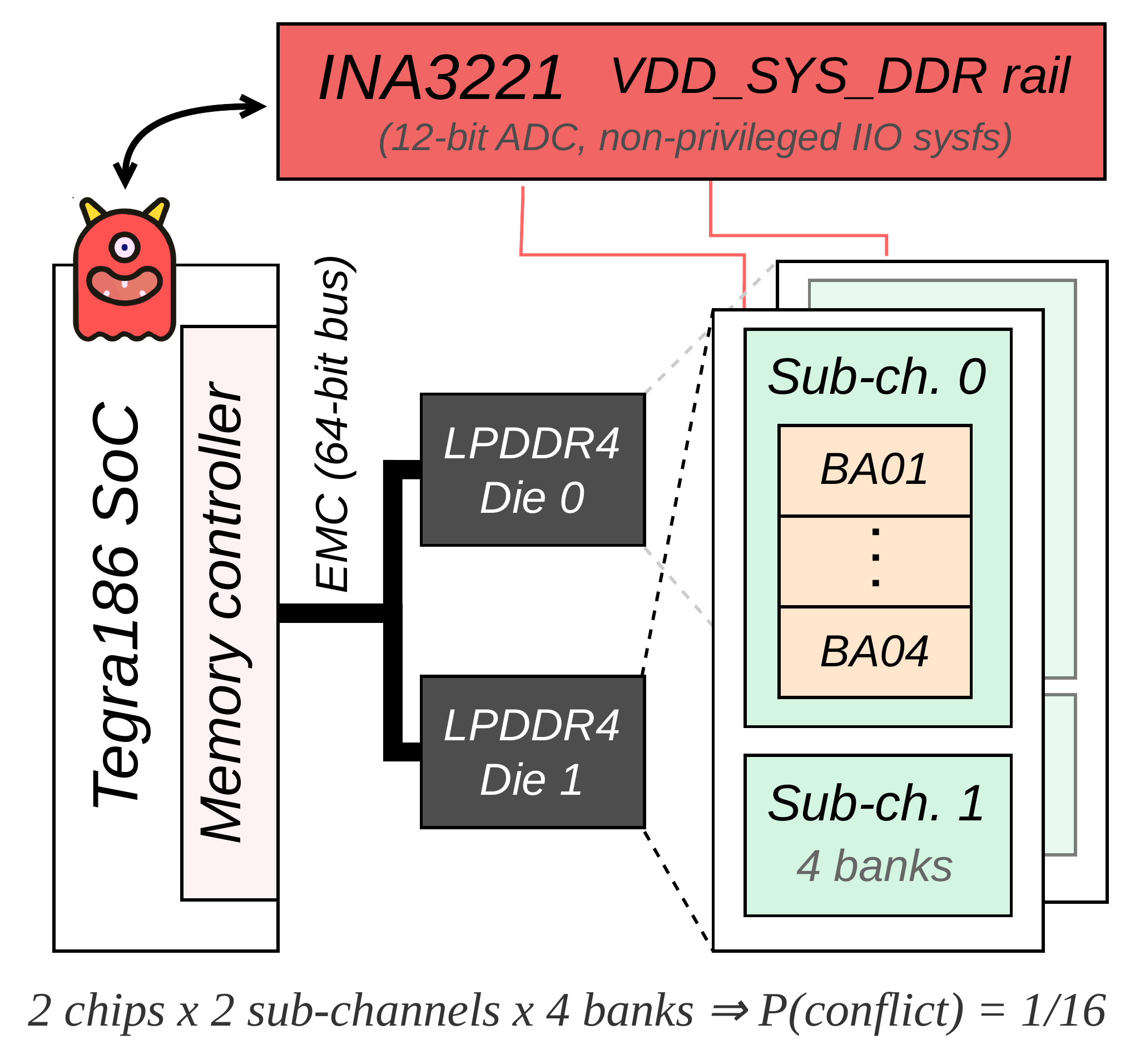}
    \caption{Jetson~TX2.}
    \label{fig:setup:tx2}
  \end{subfigure}
  \caption{Attacker setup on the Pi~5 and Jetson~TX2.}
  \label{fig:setup}
\end{figure}

\prettyparagraph{AMD ZCU102.}
A Zynq UltraScale+ MPSoC development board with a 4-bank DDR4-2400
controller and a contiguous 2-bit bank index. The DDR rail
\texttt{VCCO\_PSDDR} is monitored by a TI \texttt{INA2xx} sensor
exposed through the Linux \texttt{hwmon} interface at
$\approx\!28$\,Hz, world-readable to userspace.

\prettyparagraph{Raspberry~Pi~5.}
A Broadcom BCM2712 SoC paired with a 16-bank LPDDR4X die through an
on-package interconnect (\autoref{fig:setup:pi5}). The CPU accesses
off-chip memory through the integrated controller while the on-board
\texttt{DA9091} PMIC aggregates the DDR rail \texttt{DDR\_VDD2} and
exposes its instantaneous power consumption through the VideoCore
firmware mailbox at \texttt{/dev/vcio}, sampled at $\approx\!44$\,Hz.
Access requires membership in the standard \texttt{video} group, which
ordinary desktop users hold by default.

\prettyparagraph{NVIDIA Jetson~TX2.}
A Tegra186 SoC carrying two physically distinct LPDDR4 memory chips on
a 64-bit bonded bus (two 32-bit sub-channels) at $1331$\,MT/s, with an
aggressive DVFS policy (\autoref{fig:setup:tx2}). The DDR rail
\texttt{VDD\_SYS\_DDR} is monitored by an on-board \texttt{INA3221}
multi-channel sensor exposed through the Linux IIO interface at
$1$\,kHz with a $1$\,mA LSB.

\prettyparagraph{Adapting \textsc{Pair-Probe} per platform.}
The primitive itself is identical across platforms; what changes are
two implementation knobs chosen to keep the per-probe standard deviation
below one sensor LSB. The iteration count $N$ scales with the sensor's
LSB and noise floor: $N\approx\!1\text{--}5{\times}10^3$ on the ZCU102
and Pi~5, where the rail signal is read in mA or mW and the noise floor
is tens of $\mu$A, and $N = 50{,}000$ on the Jetson, where the
INA3221's $1$\,mA LSB requires longer integration. The Jetson also needs a $90$\,s rail warmup to suppress DVFS transients that the $1$\,kHz sensor rate would otherwise alias; the other two platforms run using their default configuration.

\prettyparagraph{Ground-truth physical addresses.}
Recovering the hierarchy mapping function in physical-address
coordinates requires the physical address of each probed virtual
address. We obtain these by querying \texttt{/proc/self/pagemap} once
per platform configuration, the same offline cost incurred by prior
DRAM reverse-engineering~\cite{pessl2016drama, plin2026knockknock}.
The online measurement loop itself requires no privileged access.

\subsection{Timing Baseline and Where It Structurally Fails}
\label{sec:dramre:timing}
We first reproduce the timing methodology of prior
work~\cite{pessl2016drama, plin2026knockknock} on all three platforms,
then evaluate the power channel under the same protocol. The timing
methodology is unchanged: for each pair $(A, B)$, the attacker times
two consecutive uncached loads, accumulates a per-pair latency
histogram, and classifies the pair by a threshold separating row-buffer
hits from conflicts. The power channel substitutes the latency measurement by power analysis; everything else in the procedure is identical.

\begin{figure*}[t]
  \centering
  \begin{subfigure}[t]{0.3\textwidth}
    \centering
    \includegraphics[width=\linewidth]{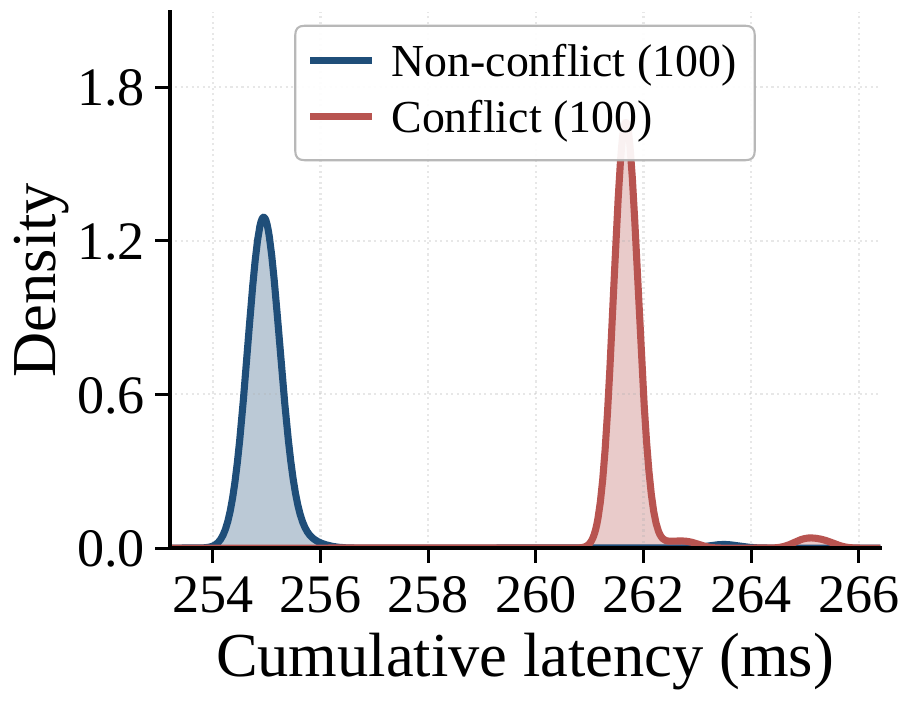}
    \caption{Timing, ZCU102.}
    \label{fig:dram:channels:a}
  \end{subfigure}\hfill
  \begin{subfigure}[t]{0.3\textwidth}
    \centering
    \includegraphics[width=\linewidth]{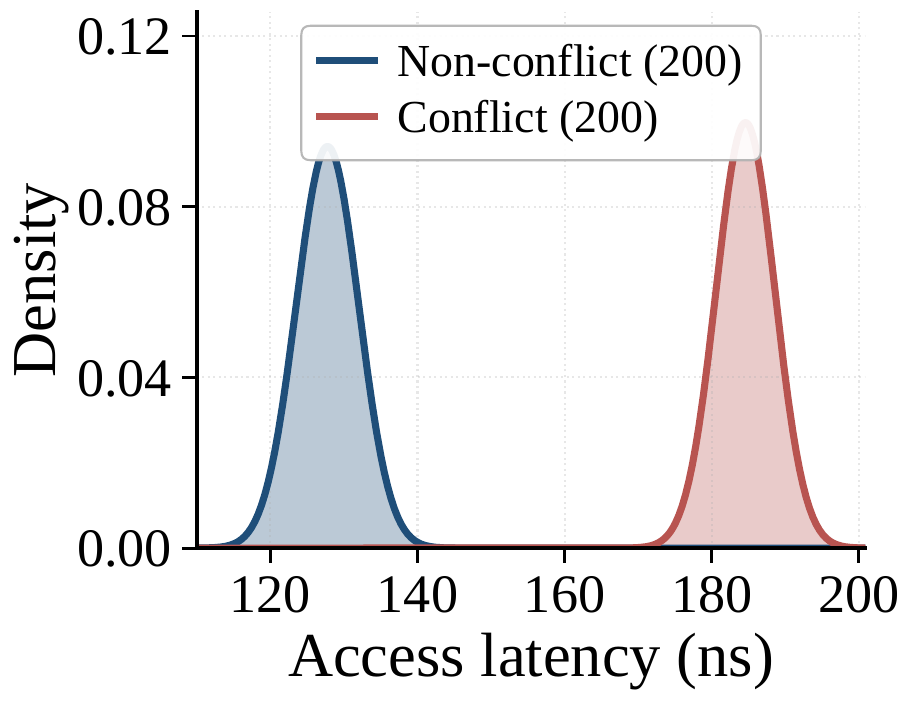}
    \caption{Timing, Raspberry~Pi~5.}
    \label{fig:dram:channels:b}
  \end{subfigure}\hfill
  \begin{subfigure}[t]{0.3\textwidth}
    \centering
    \includegraphics[width=\linewidth]{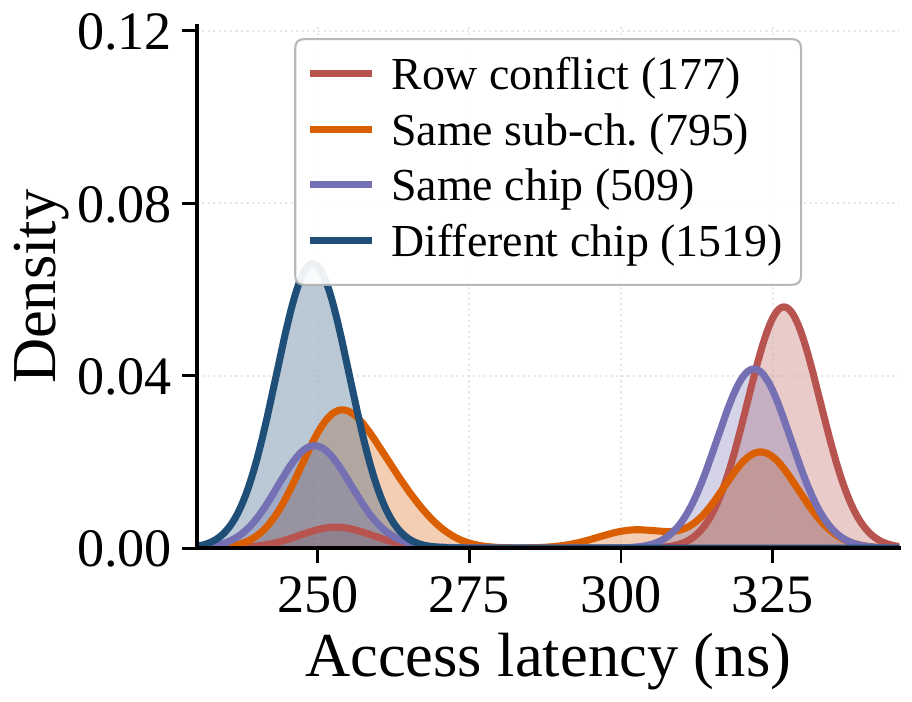}
    \caption{Timing, Jetson~TX2.}
    \label{fig:dram:channels:c}
  \end{subfigure}\\[0.5em]
  \begin{subfigure}[t]{0.3\textwidth}
    \centering
    \includegraphics[width=\linewidth]{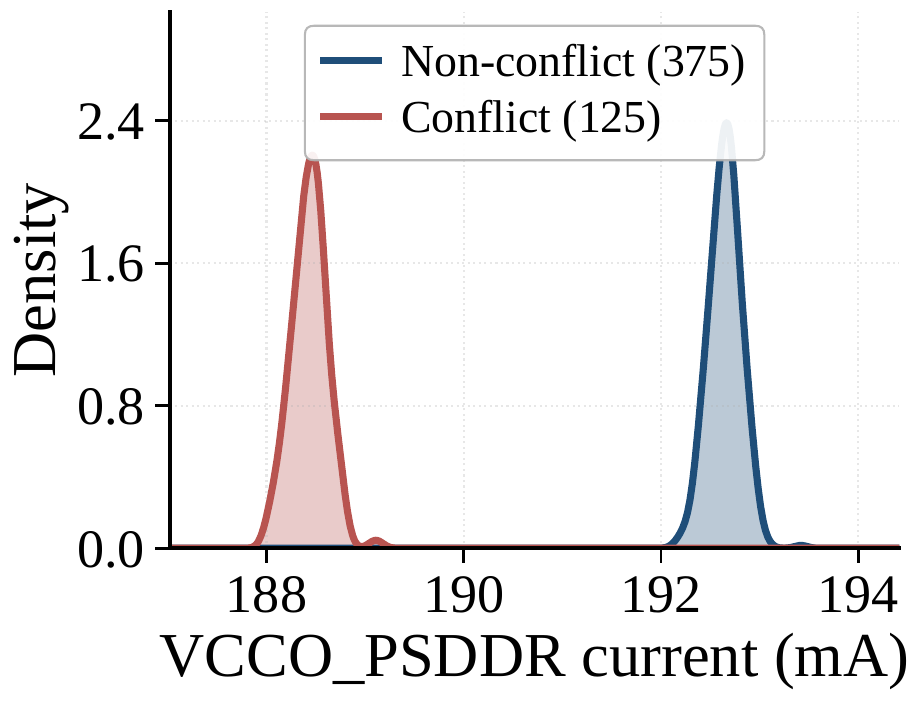}
    \caption{Power, ZCU102.}
    \label{fig:dram:channels:d}
  \end{subfigure}\hfill
  \begin{subfigure}[t]{0.3\textwidth}
    \centering
    \includegraphics[width=\linewidth]{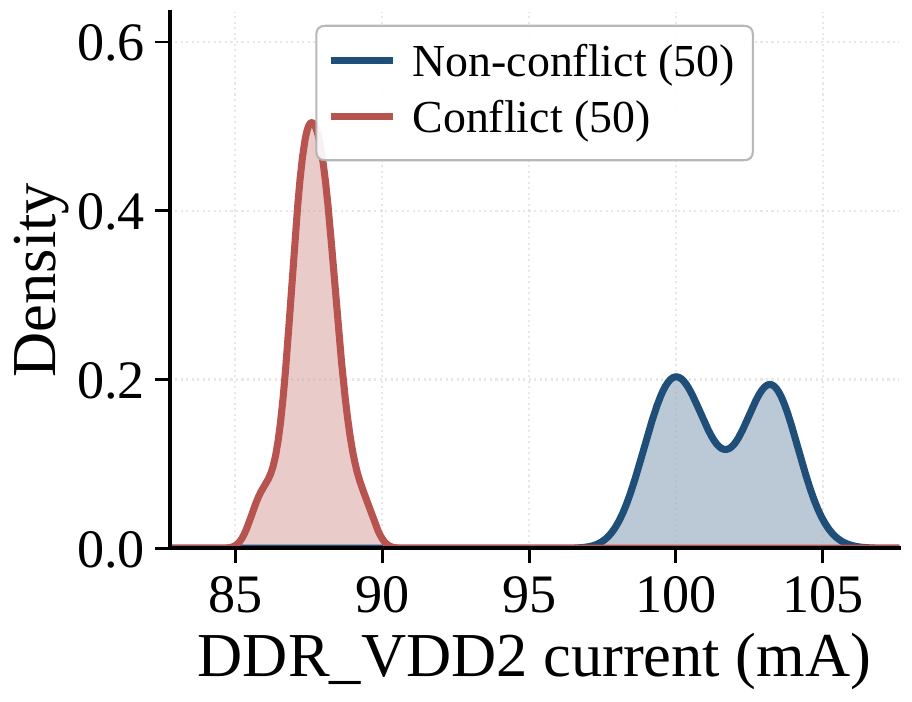}
    \caption{Power, Raspberry~Pi~5.}
    \label{fig:dram:channels:e}
  \end{subfigure}\hfill
  \begin{subfigure}[t]{0.3\textwidth}
    \centering
    \includegraphics[width=\linewidth]{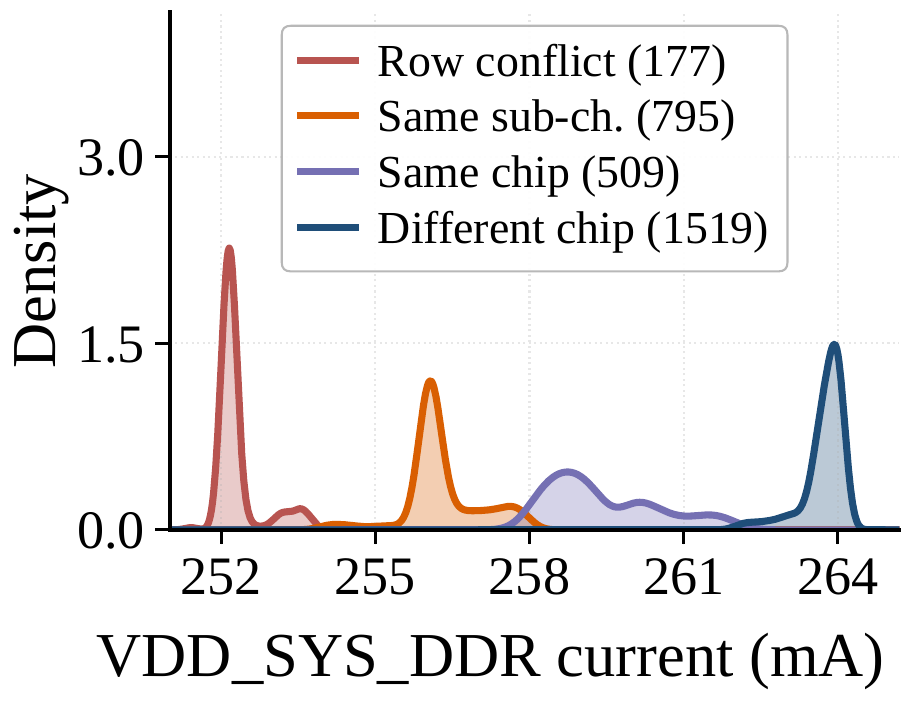}
    \caption{Power, Jetson~TX2.}
    \label{fig:dram:channels:f}
  \end{subfigure}
  \caption{Timing (top) vs.\ power (bottom) on the three evaluation
  platforms. The parenthesized number in each legend entry is the
number of probed address pairs falling in that class.}
  \label{fig:dram:channels}
\end{figure*}

The top row of \autoref{fig:dram:channels} shows the timing-channel result. On the \textbf{ZCU102} (\autoref{fig:dram:channels:a}) and the \textbf{Raspberry~Pi~5} (\autoref{fig:dram:channels:b}), the latency histograms are cleanly bimodal: row-buffer conflicts and non-conflicts form non-overlapping clusters, classifiable by a single threshold with $\mathrm{AUC} = 1.0$. On the \textbf{Jetson TX2}~(\autoref{fig:dram:channels:c}), the populations no longer separate: out of $3{,}000$ pairs, $2{,}954$ measure $249$\,ns and only $46$ measure $325$\,ns. In a 16-bank mapping, random pairs should conflict with probability $1/16$, predicting $\approx 188$ conflicts; the timer exposes a quarter of that. The information content drops to $0.11$\,bits per pair, and the bank functions cannot be uniquely recovered. This matches the limitations discussed in~\cite{plin2026knockknock}.

The bottom row of \autoref{fig:dram:channels} shows the power-channel response under the same protocol. On the \textbf{ZCU102}, $500$ pairs on the \texttt{VCCO\_PSDDR} rail form two non-overlapping clusters at $188.44$\,mA (conflicts) and $192.65$\,mA (non-conflicts) separated by a $3.05$\,mA gap, yielding $\mathrm{AUC} = 1.0$ and $100\%$ accuracy with a single threshold at $190.64$\,mA. On the \textbf{Raspberry~Pi~5}, $240$ pairs on the \texttt{DDR\_VDD2} rail split at $87.99 \pm
1.96$\,mA and $102.92 \pm 2.47$\,mA with a $6.26$\,mA gap, yielding $\mathrm{AUC} = 1.0$ and $99.6\%$ accuracy on a $500$-pair blind deployment; the same blind protocol on the ZCU102 reaches $100\%$. In both cases, the power channel matches the timing channel: a single scalar threshold suffices.

On the Jetson, the power channel exposes a structure that timing cannot distinguish. The measurements form four non-overlapping peaks at $252.2$, $256.2$, $259.2$, and $264.0$\,mA, with intra-class spread $\sigma \le 0.67$\,mA and adjacent gaps of at least $3.0$\,mA. The four levels correspond to four DRAM relations: same bank, same sub-channel, same chip, and distinct chip. Those levels instantiate the multi-level oracle of \autoref{sec:dramre:recovery} with $r = 3$ and rank increments $(q_1, q_2, q_3) = (1, 1, 2)$. The correspondence between power levels and hierarchy classes is not assumed, but it will be validated by the staged recovery of \autoref{sec:dramre:recovery-results}, where any other permutation of labels fails to produce a consistent null-space basis. Two further consistency checks support the labeling. First, the chip-sharing rate over $3{,}000$ random pairs is $50.6\%$, matching the uniform distribution over two chips. Second, across $10$ repetitions per pair, $99.3\%$ of probes land in the same of the four bins, confirming per-probe stability of the labels. We quantify the resulting information content as $2.88$\,bits per pair, against $0.11$\,bits for the timing channel on the same platform.

\subsection{Reverse Engineering beyond Timing}
\label{sec:dramre:recovery-results}
On the two platforms where the power channel is bimodal
(\autoref{fig:dram:channels:d},~\subref{fig:dram:channels:e}), the
recovery reduces to the standard null-space procedure of prior
work~\cite{plin2026knockknock}. 
On the \textbf{ZCU102}, $20{,}000$ pairs yield $4{,}980$ conflicts ($24.9\%$, matching the $1/4$ rate of a 4-bank system) and the null space recovers the contiguous 2-bit bank index. On the \textbf{Raspberry~Pi~5}, $500$ pairs yield $31$ conflicts ($6.2\%$, matching $1/16$ for 16 banks) and recovery gives a 16-bank mapping. Wall-clock cost is dominated by measurement ($\approx 500$\,s on the ZCU102, $\approx 30$\,s on the Pi~5); computation is milliseconds. The full recovered mapping functions are listed in Appendix~\ref{appendix:mapping}.

On the \textbf{Jetson~TX2}, the four-level power signal allows us to perform richer recovery than the binary distribution model. The binary same-bank oracle of prior work would wait for $1/16$ rare row-conflict pairs:
\begin{align*}
\Pr[\mathrm{same\ bank}]
  ={}& \Pr[c(D)=0] \\
  &{}\cdot \Pr[s(D)=0 \mid c(D)=0] \\
  &{}\cdot \Pr[f_1(D)=f_2(D)=0 \\
  &{}\hphantom{\cdot{}\Pr[}\mid c(D)=s(D)=0] \\
  ={}& \tfrac{1}{2} \cdot \tfrac{1}{2} \cdot \tfrac{1}{4}
   = \tfrac{1}{16},
\end{align*}
paying a $16\times$ collision penalty on every probed row before obtaining one useful constraint. Our multi-level oracle exposes the intermediate equalities and recovers the hierarchy sequentially, from the most probable equality to the least probable.

\prettyparagraph{Stage~1: chip selector.}
All pairs with $L_i \ne \mathrm{distinct}$ satisfy $c(D_i) = 0$. We
collect $D_c = \{D_i : L_i \ne \mathrm{distinct}\}$ and solve $D_c
x^\top = 0$; at rank $n-1$, $\mathrm{nullspace}(D_c) = \mathrm{span}(c)$.
Since $\Pr[c(D) = 0] = 1/2$, this stage requires
\[
N_c \ge \frac{2}{1 - \theta_c} \log_2\!\left(\frac{2^{n-1}-1}{\varepsilon_c}\right)
\]
measurements in expectation.

\prettyparagraph{Stage~2: sub-channel selector.}
Knowing $c$, we restrict the search to $\ker(c) = \{D : c(D) = 0\}$.
Inside this halved space, same-sub-channel pairs satisfy $s(D) = 0$.
We collect $D_{c,s} = \{D_i : c(D_i) = 0,\ L_i \in \{\mathrm{subch},
\mathrm{conflict}\}\}$ and solve $D_{c,s} x^\top = 0$; at full rank,
$\mathrm{nullspace}(D_{c,s}) = \mathrm{span}(c, s)$, and $s$ is the
independent direction not already in $\mathrm{span}(c)$. Since
$\Pr[s(D) = 0 \mid c(D) = 0] = 1/2$, the expected cost is
\[
N_s \ge \frac{2}{1 - \theta_s} \log_2\!\left(\frac{2^{n-2}-1}{\varepsilon_s}\right).
\]

\prettyparagraph{Stage~3: bank selectors.}
Knowing $c$ and $s$, we sample inside $\ker(c, s) = \{D : c(D) = s(D)
= 0\}$, a space already reduced to one quarter of the address space.
Within this subspace, row-conflict pairs satisfy $f_1(D) = f_2(D) = 0$
with probability $1/4$, four times denser than the $1/16$ rate the
binary oracle would observe in the full space. We collect $D_{c,s,f}$
and solve for the null space, recovering $\mathrm{span}(c, s, f_1,
f_2)$; the bank-selector space is the complement of $\mathrm{span}(c,
s)$ inside this span. The expected cost is
\[
N_f \ge \frac{4}{1 - \theta_f} \log_2\!\left(\frac{2^{n-4}-1}{\varepsilon_f}\right).
\]

\prettyparagraph{Summary on TX2.}
The hierarchy-aware procedure replaces a single rare $1/16$ event with
three conditional events of probabilities $1/2$, $1/2$, and $1/4$,
each sampled inside the kernel recovered at the previous stage. In the
language of \autoref{thm:dramre:bound}, the rank increments are
$(q_1, q_2, q_3) = (1, 1, 2)$, the binary collision factor is
$2^{1+1+2} = 16$, and the hierarchy-aware factor is
$2^1 + 2^1 + 2^2 = 8$. Thus, the per-trace improvement is
$2\times$ before accounting for the stage-wise sampling restriction; accounting for it, the bank stage alone is four times denser
than in the full address space. The near-zero up-leak ($\theta_j
\approx 0$) assumed by the bound holds as the $\ge 3$\,mA
inter-class gap in \autoref{fig:dram:channels:f} dwarfs the
$0.67$\,mA intra-class spread, making label-flip errors vanishingly
rare.

\begin{figure}[h]
  \centering
  \includegraphics[width=\linewidth]{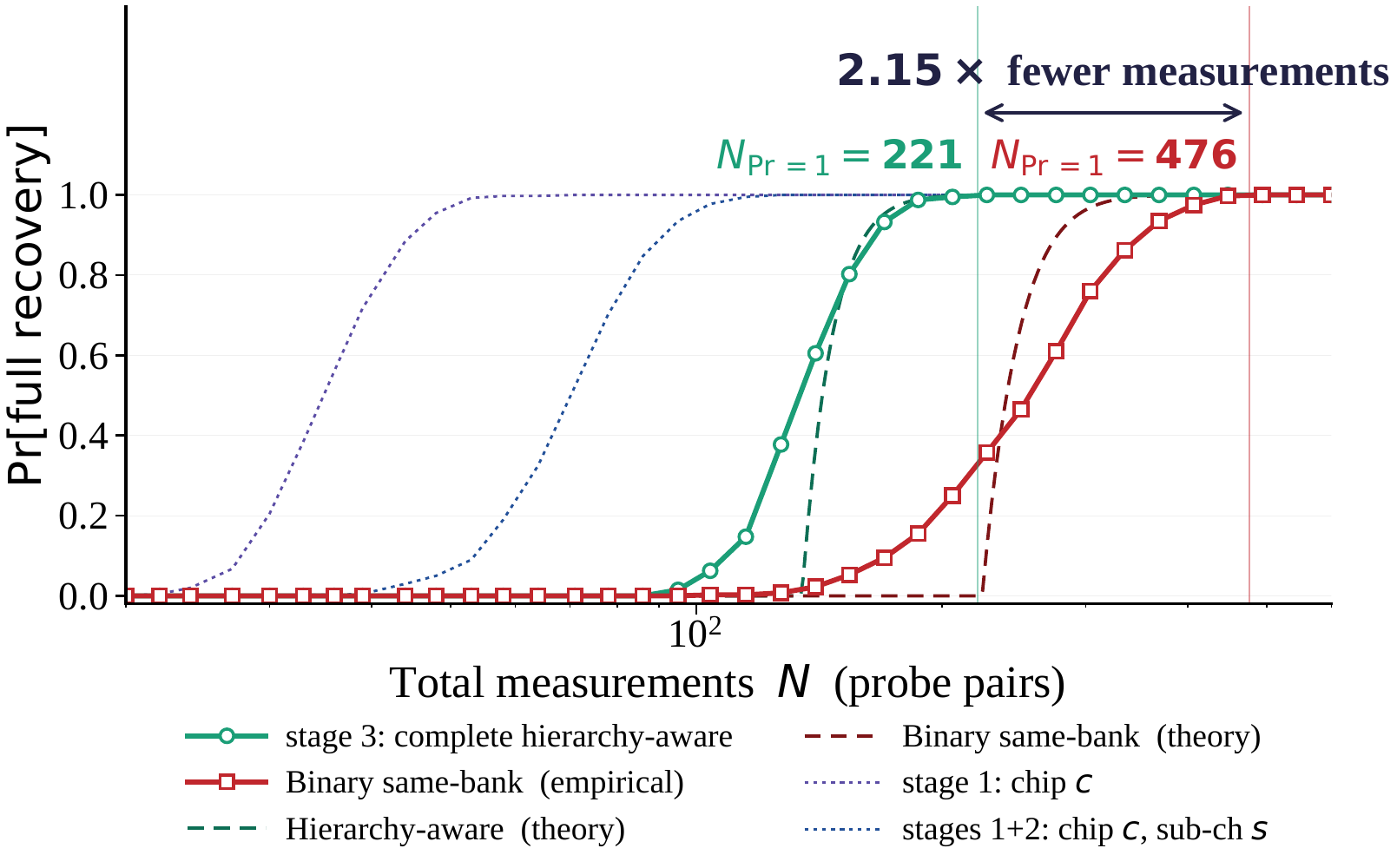}
  \caption{Recovery probability vs.\ measurements (Jetson~TX2).}
  \label{fig:dram:recovery}
\end{figure}

\autoref{fig:dram:recovery} validates the procedure on real hardware.
The binary oracle waits for $1/16$ rare row-conflict pairs and only
completes at $N_{\Pr=1} = 476$. The hierarchy-aware procedure spends
its first measurements on the cheap chip and sub-channel stages
(probabilities $1/2$ and $1/2$), then samples inside $\ker(c, s)$
where conflicts are four times denser, and completes at $N_{\Pr=1} =
221$. This is a $2.15\times$ reduction in total measurements for the
same end-to-end recovery, and the empirical curves track the
theoretical bound tightly. The Jetson result therefore demonstrates
the multi-level methodology of \autoref{sec:dramre:recovery} on real
hardware in the regime it was designed for: a channel exposing $r > 1$
hierarchy levels, where binary recovery is strictly suboptimal.

\section{\sysname-DATA: DRAM Data Recovery}
\label{sec:dramhw}
\sysname-RE recovered \emph{where} memory lives. \sysname-DATA
instantiates the same primitive (\autoref{sec:channel:wfp}) with the
data-dependent stimulus $\mathcal{H}_{\mathrm{D}}$, recovering \emph{what} is moving over the data bus. A DRAM read draws a rail current with two data-dependent terms~\cite{ghose2018your}: a static term set by the line's Hamming weight, and a dynamic term set by the bits that toggle on the bus between two consecutive transfers. A single kernel gadget, described in~\autoref{lst:gadget:ref}) in Appendix~\ref{sec:appendix:gadgets}, exploits both, and we also evaluate it under speculative execution.

\subsection{Setup}
\label{sec:dramhw:setup}
We run all experiments on the Pi~5. 
Of the rails the PMIC exposes, two carry the DRAM signal: \texttt{DDR\_VDD2\_A},
the $\approx\!1.1$\,V core-array supply, and \texttt{DDR\_VDDQ\_A}, the
$\approx\!0.6$\,V DQ I/O supply. The other ten stay at their noise floor
and serve as negative controls.

\begin{figure}[t]
  \centering
  \begin{subfigure}[t]{0.49\linewidth}
    \centering
    \includegraphics[width=\linewidth]{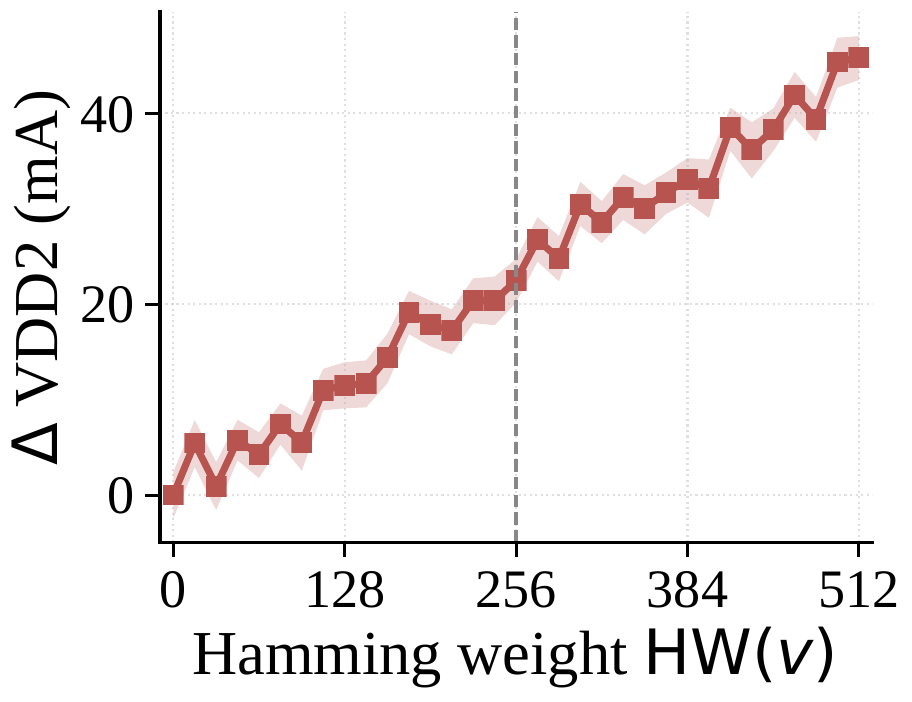}
    \caption{\texttt{DDR\_VDD2\_A} (core rail).}
    \label{fig:dramhw:rails:vdd2}
  \end{subfigure}
  \begin{subfigure}[t]{0.49\linewidth}
    \centering
    \includegraphics[width=\linewidth]{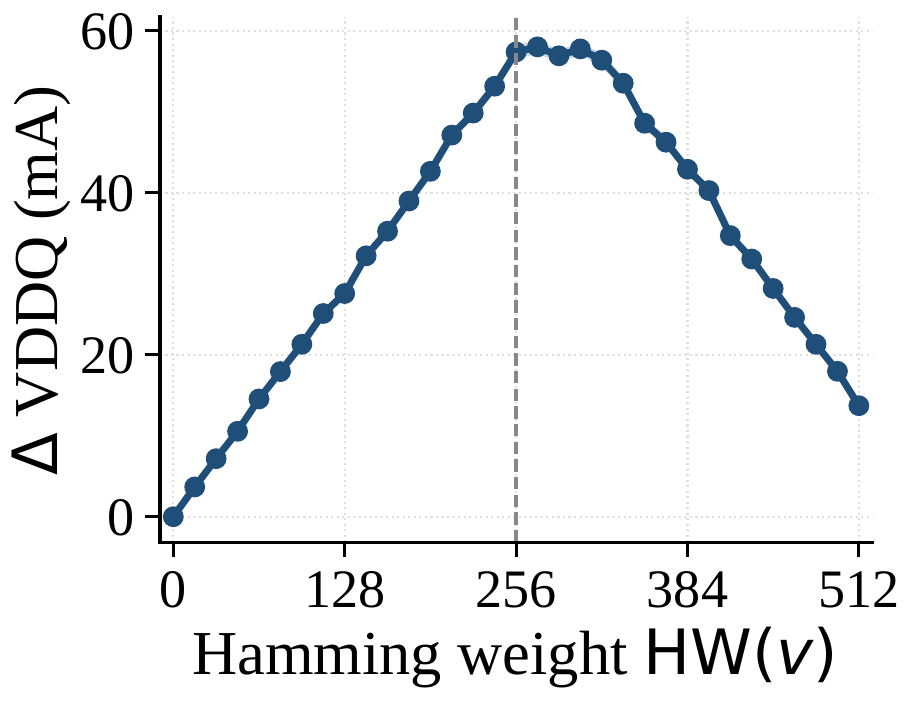}
    \caption{\texttt{DDR\_VDDQ\_A} (I/O rail).}
    \label{fig:dramhw:rails:vddq}
  \end{subfigure}\hfill
  \caption{Hamming-weight response of the two DRAM rails. Both plots use
    amplification: the target line is replicated across contiguous cache
    lines and read repeatedly per trigger.}
  \label{fig:dramhw:rails}
\end{figure}

\autoref{fig:dramhw:rails} shows how the two rails respond as the line's
Hamming weight sweeps from $0$ to $512$. The core rail
\texttt{DDR\_VDD2\_A} (\autoref{fig:dramhw:rails:vdd2}) rises linearly
with weight ($R^2 = 0.98$): each set bit adds a fixed current. The I/O rail \texttt{DDR\_VDDQ\_A} (\autoref{fig:dramhw:rails:vddq}) traces
an inverted V instead: current climbs to $\mathrm{HW}\approx256$, then
falls toward all-ones. This fold is symmetric, so each VDDQ level maps to
two weights. This matches JEDEC LPDDR4 Data Bus Inversion (DBI)~\cite{jedec2014lpddr4}. The fold appears on the bus
rail and not the core rail, because DBI inverts bus bits while the array
drives the un-inverted data.

\subsection{Shared primitive}
\label{sec:dramhw:primitive}
Both attacks (\autoref{sec:dramhw:hw} and \autoref{sec:dramhw:toggle}) instantiate the \sysname primitive
(\autoref{sec:channel:wfp}) with a data-dependent stimulus
$\mathcal{H}_{\mathrm{D}}(G,V)$. For those attacks we consider two memory blocks, $V$ and $G$, in the same DRAM row at adjacent columns, the placement that maximizes
the signal-to-noise ratio (SNR)~\cite{ghose2018your}: the two reads hit the open row and the same Data Input/Output (DQ) lanes, so the $G\!\rightarrow\!V$ transition drives the full per-lane toggle onto the rails with no intervening activation (ACT). As discussed in \autoref{sec:disc:loc}, this placement is not a strict requirement, however. When $G$ and $V$ do not share a row, for instance when they sit in different banks, the signal remains exploitable. $V$ holds the unknown 64-byte victim content $v=(v_1,\dots,v_{64})$; $G$ holds an attacker-controlled block $g=(g_1,\dots,g_{64})$. The stimulus flushes both and reads $G$ then $V$. Sharing the open row and the same DQ lanes, the two reads drive $g$ then $v$ over the bus. Run for $N$ iterations, the primitive returns the scalar $\mu$ of \autoref{alg:primitive}.

The $G\!\rightarrow\!V$ transition toggles the bits in which $g$ and $v$
differ, so the dynamic term scales with Hamming distance,
\begin{equation}
  \mathrm{HD}(G,V)
  \;=\; \mathrm{HW}(g \oplus v)
  \;=\; \sum_{j=1}^{64}\mathrm{HW}(g_j \oplus v_j),
  \label{eq:primitive:hd}
\end{equation}
a sum of 64 independent per-byte terms. The reading is affine in this
distance,
\begin{equation}
  \mu \;=\; \alpha + \gamma\,\mathrm{HD}(G,V) + \varepsilon,
  \label{eq:primitive:model}
\end{equation}
with per-bit current $\gamma$, baseline $\alpha$, and noise $\varepsilon$
that averages out over $N$. Equations~\ref{eq:primitive:hd}
and~\ref{eq:primitive:model} hold throughout the remainder of this section. The choice of $G$ sets the purpose:



\begin{itemize}
    \item \textbf{Hamming-weight oracle} ($G=\mathbf{0}$):
    An attacker's memory block full of 0s collapses the distance to the victim's block Hamming weight (\autoref{sec:dramhw:hw}):
    \begin{equation}
      \mu \;=\; \mathrm{HD}(\mathbf{0},V) \;=\; \mathrm{HW}(v),
      \label{eq:primitive:hw}
    \end{equation}
    
    \item \textbf{Data value recovery} ($G$ random):
    The memory block is set with a uniform-random $g_i$ per trace, and then the attacker drives an unprofiled attack (\autoref{sec:dramhw:toggle}). 
\end{itemize}

\subsection{Hamming-Weight Recovery}
\label{sec:dramhw:hw}
We run the primitive in its $G=\mathbf{0}$ mode, where $\mu$ reads out $\mathrm{HW}(v)$. Recovery is profiled: we first build a template offline and then match against it online.

Offline, the template we build maps a rail reading to a Hamming
weight. We want both VDDQ and VDD2 rails to drive the estimate, because each is informative over a different part of the range.
VDDQ is steep, and therefore discriminative.
However, due to the DBI creating the measurement symmetry shown in \autoref{fig:dramhw:rails:vddq}, each actual sample  might correspond to two different values: $k$ and $512-k$.
Because of this, it is informative for only half of the range.
In contrast, VDD2 is monotonic across the entire range but offers less discrimination between values.

\begin{figure}[t]
  \centering
  \begin{subfigure}[t]{0.49\linewidth}
    \centering
    \includegraphics[width=\linewidth]{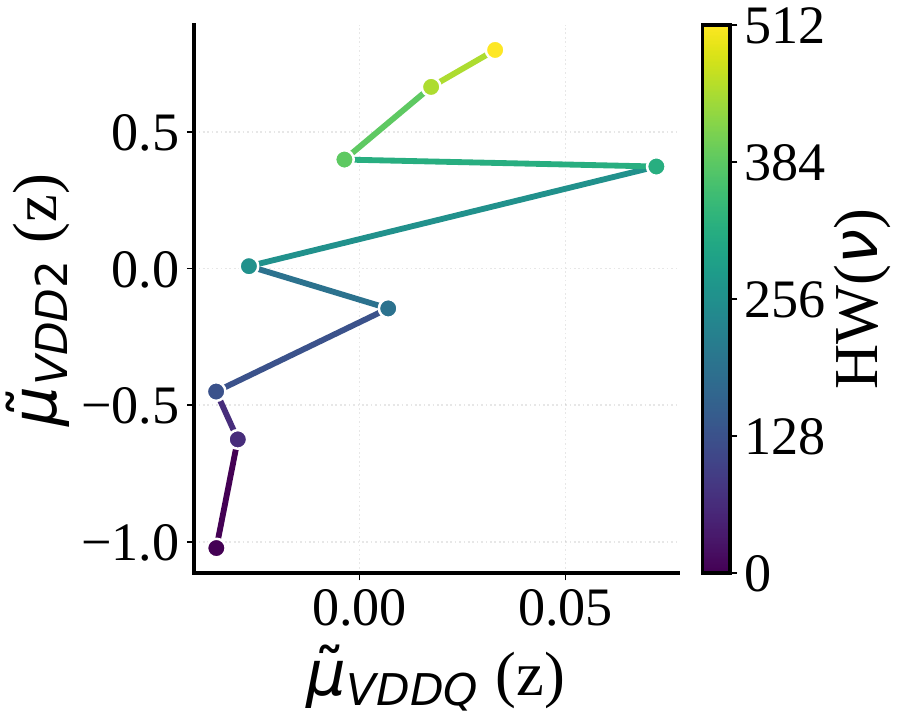}
    \caption{Offline joint template.}
    \label{fig:dramnarch:jointtemp}
  \end{subfigure}
  \begin{subfigure}[t]{0.49\linewidth}
    \centering
    \includegraphics[width=\linewidth]{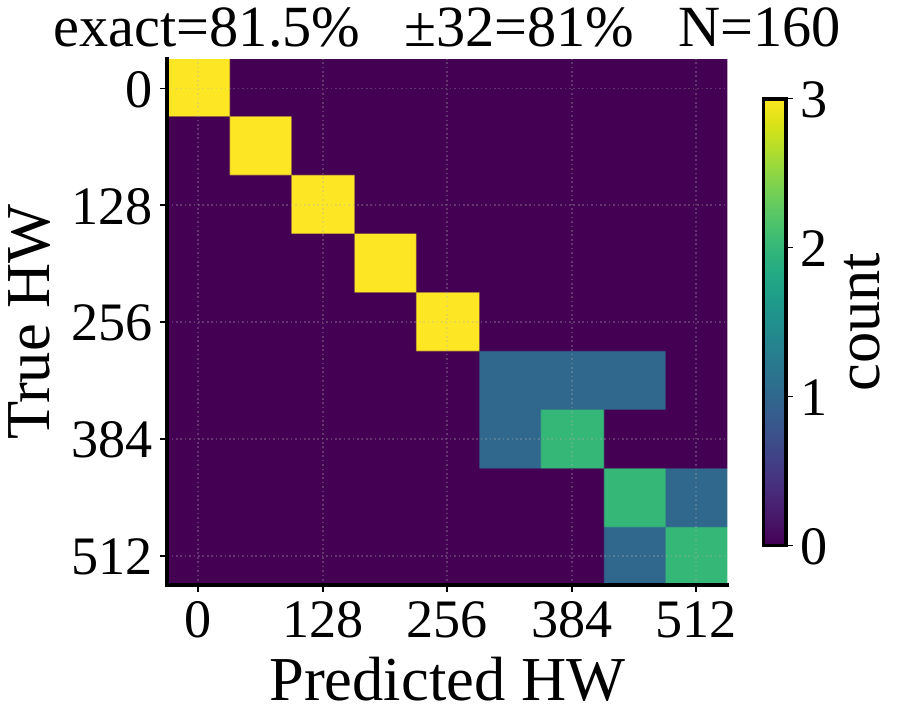}
    \caption{Confusion matrix.}
    \label{fig:dramarch:cm}
  \end{subfigure}
  \caption{Hamming-weight recovery in the $G=\mathbf{0}$ mode: (a) the
  offline template, (b) the online per-line recovery.}
  \label{fig:dramnarch:hw}
\end{figure}

We resolve this geometrically. We plot one rail against the other, so each
Hamming weight $n$ becomes a single point
$\widetilde{v}_n=(\widetilde{\mu}_{\mathrm{VDDQ}}(n),\,
\widetilde{\mu}_{\mathrm{VDD2}}(n))$; each rail is first z-scored (centered
and scaled to unit variance) so the two axes are comparable. As $n$ runs
from $0$ to $512$, the points trace the calibration curve of
\autoref{fig:dramnarch:jointtemp}. We then identify each reading by how far
along this curve it lies, i.e., we calculate its arc length from $n=0$:
\begin{equation}
  s(n_i)=\sum_{j<i}\|\widetilde{v}_{j+1}-\widetilde{v}_j\|_2 .
  \label{eq:arclength}
\end{equation}
The arc length thus combines both rails into a single monotonic coordinate, using VDDQ and VDD2 to break the fold-symmetry ambiguity and keeping VDDQ's high separability. Illustrated in \autoref{fig:dramnarch:jointtemp}, this joint template is more accurate than either rail alone.

Online, the primitive returns the rail pair
$(\mu_{\mathrm{VDDQ}},\mu_{\mathrm{VDD2}})$ for the unknown line. We
z-score it, project it onto the template polyline
(\autoref{fig:dramnarch:jointtemp}), and read the arc-length coordinate
of the nearest point. Linear interpolation between the bracketing
vertices inverts it to $\widehat{\mathrm{HW}}=s^{-1}(\mu)$, unambiguous
across the fold.

\autoref{fig:dramarch:cm} reports the online phase. At $N=450$ traces per
line ($\approx\!150$\,s), per-line recovery over the full $0$--$512$ range
reaches a mean absolute error of $5.1$ bits. The static term is therefore a
full-range Hamming-weight oracle: this unprivileged rail sampling estimates
a victim line's weight to within a few bits.

\subsection{Data Value Recovery}
\label{sec:dramhw:toggle}
Hamming-weight recovery returns only how many bits a byte sets, and many
bytes set the same number of bits (\texttt{0x01}, \texttt{0x02}, and
\texttt{0x80} all have weight one). The weight therefore identifies a class
of values, not a single one. The dynamic term resolves the value: the
Hamming distance $\mathrm{HW}(g\oplus v_j)$ to an attacker-chosen byte $g$
depends on \emph{which} bits differ, not just on how many $v_j$ sets.
Sweeping $g$ over many values leaves only the true $v_j$ consistent with
every measured distance.
We therefore run the primitive in its random-$G$ mode
(\autoref{sec:dramhw:primitive}) and apply per-byte CPA to recover the exact value.

\begin{figure}[t]
  \centering
  \includegraphics[width=0.65\linewidth]{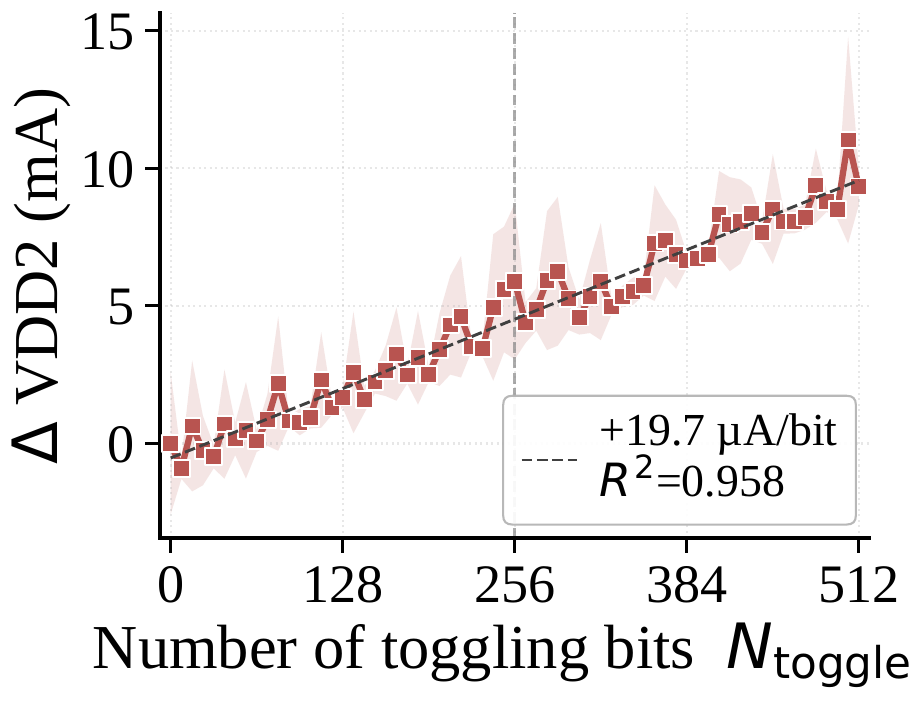}
  \caption{Linear HD response of \texttt{DDR\_VDD2\_A} to full-line bit
    toggling. The horizontal axis $N_{\mathrm{toggle}}$ is the number of
    toggling bits, i.e. the Hamming distance $\mathrm{HD}(G,V)$ of \autoref{eq:primitive:hd}}
  \label{fig:dramhd:toggle}
\end{figure}

Recovery targets VDD2. \autoref{fig:dramhd:toggle} sweeps
$\mathrm{HD}(G,V)$ from $0$ to $512$ at fixed $\mathrm{HW}(G)$: VDD2 rises
linearly, $\approx\!20\,\mu$A per toggled bit ($R^2=0.96$), so the rail
reads out $\mathrm{HD}(G,V)$ directly. We observe that VDDQ shows no toggling response above the sample noise, consistent with the low I/O-supply SNR reported by~\cite{ghose2018your}.

Each trace draws a random $g_i$ and records $\mu_i$, which grows linearly with the number of toggling bits (i.e., $\mathrm{HD}(g_i,v)$), as the sum of the independent per-byte distances.
As a consequence, per-byte CPA could theoretically recover each $v_j$ separately: for every byte position, it correlates the readings against the toggle count predicted by each candidate value and keeps the best-correlated one as the potentially correct byte.
Two details stabilize the run: every $16$th trace uses $G=\mathbf{0}$ as a drift anchor subtracted before CPA, and traces whose reading lies more than three median absolute deviations (MAD) from the local median are dropped.
We declare a byte recovered when its true value reaches rank~$1$ among the $256$ candidates.

\begin{figure}[t]
  \centering
  \includegraphics[width=0.9\linewidth]{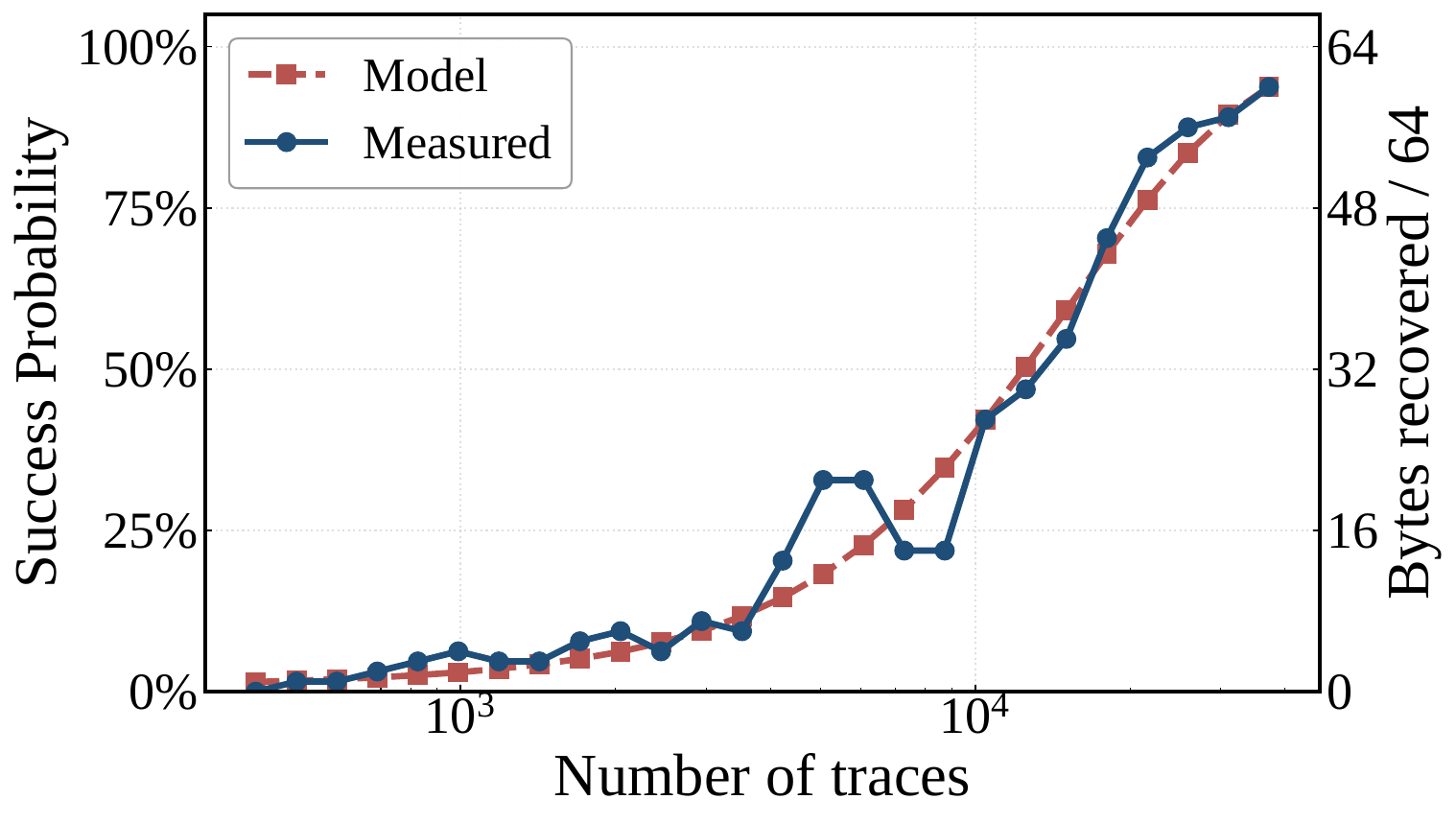}
  \caption{Blind byte recovery with CPA.}
  \label{fig:dramhd:success}
\end{figure}

We evaluate against a 64-byte secret that spans all Hamming-weight
classes, including the extremes \texttt{0x00} and \texttt{0xFF}. The
attack is blind: CPA only sees the pairs $(g_i,\mu_i)$, and the ground
truth is only used for evaluation.
\autoref{fig:dramhd:success} shows the per-byte success rate against
trace count; the noise floor decays as $1/\sqrt{N}$. At $N\approx64.5$k
traces ($\approx\!19$\,h), the CPA recovers $63/64$ bytes at
rank~$1$, with the last byte at rank~$2$. The secret therefore lies in a
2-candidate space (guessing entropy $=1$\,bit), settled by one
verification query. The weight extremes recover cleanly: the all-zero and
all-one bytes, the least-variant lanes, both reach rank~$1$. Overall, we can conclude that recovery is value-agnostic.

We run two verifications. The first confirms the leakage is real: we re-pair each $g_i$ with a $\mu_i$ from the wrong trace and rerun the CPA, and recovery drops to $0/64$. The signal therefore comes from the data-dependent $G\!\rightarrow\!V$ transfer, not a measurement artifact. The second explains the one missed byte: its true value trails a competitor by a single bit, and the gap does not shrink with more traces. This is harmless, since the secret already lies in the 2-candidate space that a single verification query resolves.

\subsection{Recovery Under Speculative Execution}
\label{sec:dramhw:spectre}
Subsections \ref{sec:dramhw:hw} and \ref{sec:dramhw:toggle} have shown how to recover Hamming weight and data by reading lines that a \emph{committed} instruction loads. We now show how both modes still work when the lines are speculatively fetched only on a mispredicted path, as in Spectre~\cite{kocher2020spectre,kogler2023collidepower}.
This implies that the read of the victim block $V$ never updates the architectural state, making \sysname{}'s threat model strictly stronger, and broadening the set of exploitable gadgets.

\begin{figure}[t]
\centering
\begin{tikzpicture}[
  >=Stealth, node distance=0.4cm,
  stage/.style={draw, rounded corners=2pt, minimum height=1.0cm,
                text width=1.7cm, align=center, font=\small,
                inner sep=2pt},
  probe/.style={draw=orange!80!black, rounded corners=2pt,
                text width=3.5cm, align=center, font=\small,
                inner sep=3pt}
]
\node (s1) [stage]
      {\textbf{1.\,Train}\\[1pt]\texttt{bl} pushes \texttt{3:} on RSB};
\node (s2) [stage, right=of s1]
      {\textbf{2.\,Window}\\[1pt]small latency holds \texttt{ret}};
\node (s3) [stage, right=of s2, draw=red!70!black, dashed]
      {\textbf{3.\,Speculate}\\[1pt]linefill burst $G\!\to\!V$ at \texttt{3:}};
\node (s4) [stage, right=of s3, draw=green!50!black]
      {\textbf{4.\,Commit}\\[1pt]\texttt{ret} resolves, squash, run \texttt{4:}};
\node (p) [probe, below=0.75cm of s3]
      {\textbf{Rail probe}\\[1pt]
       $\Delta I_{\mathrm{VDD2}}\!\approx\!\gamma_2\,\mathrm{HD}(G, V)$};
\draw[->] (s1) -- (s2);
\draw[->, dashed, red!70!black] (s2) -- (s3);
\draw[->] (s3) -- (s4);
\draw[->, orange!80!black] (s3) -- (p);
\end{tikzpicture}
\caption{Spectre-RSB gadget for \sysname-DATA.}
\label{fig:dramhw:spectre-flow}
\end{figure}
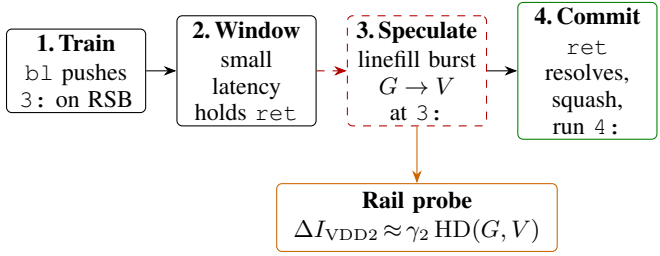

\autoref{fig:dramhw:spectre-flow} shows the gadget (\autoref{lst:gadget:spec}). A \texttt{bl} pushes the address of
label \texttt{3:} onto the return-stack buffer
(RSB)~\cite{koruyeh2018spectre}, training it to predict \texttt{3:} as
the return target. A short-latency operation produces the \texttt{ret}
target, so the \texttt{ret} resolves late. During that window the RSB
runs the instruction at \texttt{3:} speculatively and issues the linefills of $G$ then $V$,
producing one $G\!\rightarrow\!V$ bus transition. Both lines are flushed
beforehand, so each speculative load misses and triggers a DRAM linefill.
When the \texttt{ret} resolves, the path commits to \texttt{4:} and
squashes the burst, but the linefills it already dispatched complete on
the rail.

The window only has to cover the load dispatch, a few cycles, not the full DRAM round-trip of hundreds of cycles: a dispatched linefill becomes an outstanding memory transaction. A short window therefore maximizes the
speculative trigger rate. Any operation that delays the \texttt{ret} long
enough serves as the latency source.

\subsubsection*{Hamming-weight recovery ($G=\mathbf{0}$)}
\begin{figure}[t]
  \centering
  \begin{subfigure}[t]{0.49\linewidth}
    \centering
    \includegraphics[width=\linewidth]{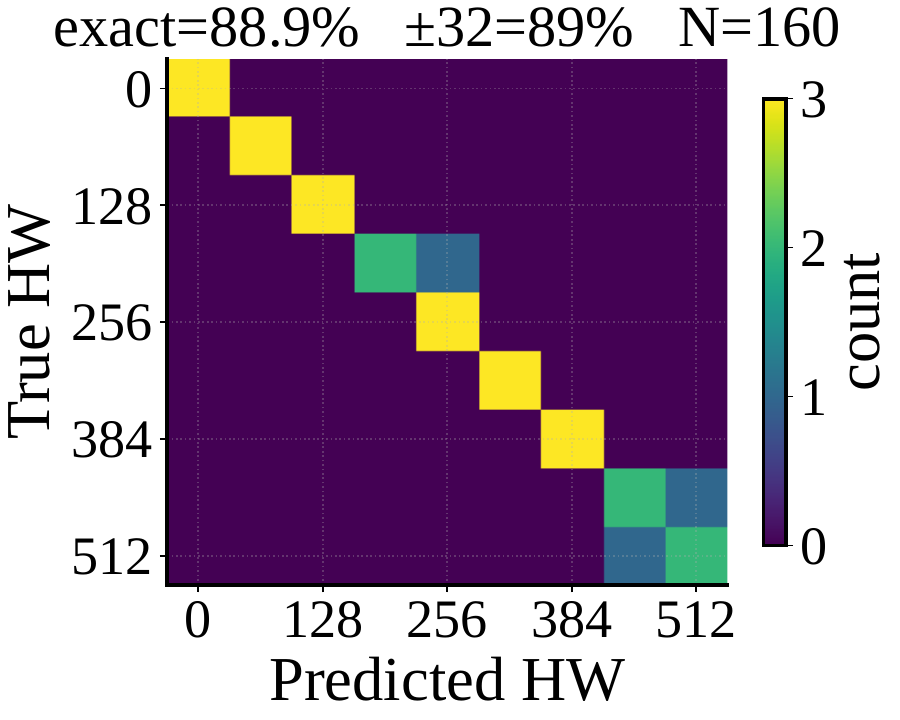}
    \caption{Hamming-weight recovery.}
    \label{fig:dramhw:spectre:hw}
  \end{subfigure}\hfill
  \begin{subfigure}[t]{0.49\linewidth}
    \centering
    \includegraphics[width=\linewidth]{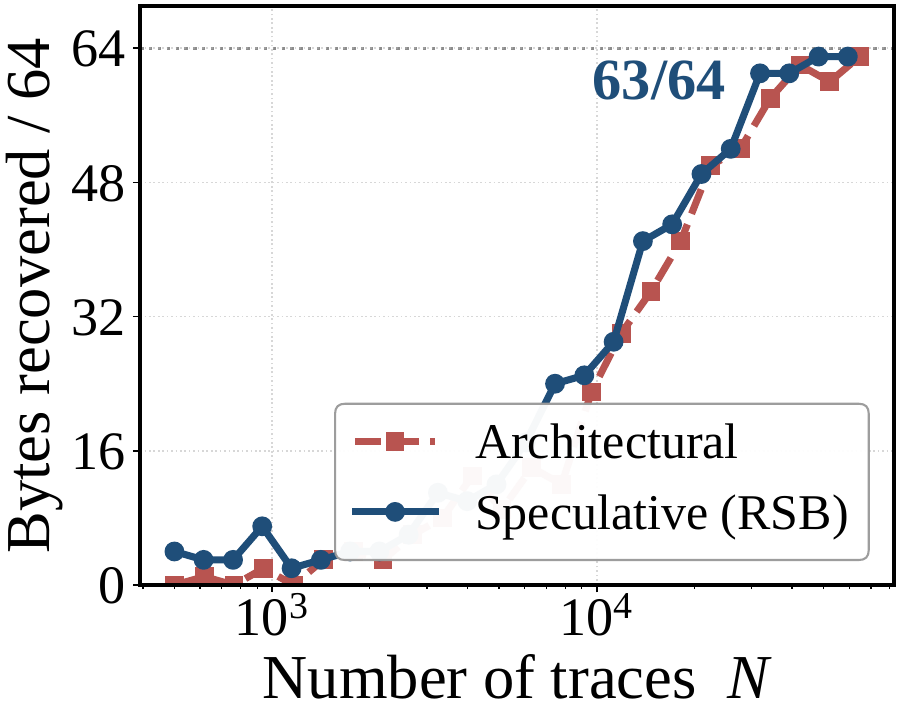}
    \caption{Value recovery.}
    \label{fig:dramhw:spectre:cpa}
  \end{subfigure}
  \caption{Speculative (RSB) recovery of an unknown $64$-byte kernel
    block.}
  \label{fig:dramhw:spec}
\end{figure}

With $G=\mathbf{0}$ the transition reduces to $\mathrm{HW}(V)$, so the
template of \autoref{sec:dramhw:hw} recovers the Hamming weight of a line read
only on the squashed path. It reaches $\approx\!89\%$ accuracy over the nine
weight classes, against an $11\%$ uniform random baseline in \autoref{fig:dramhw:spectre:hw}.

\subsubsection*{Value recovery (random $G$)}
We run the same per-byte CPA under speculation. The window length is the lever. A minimal $\approx\!2$-cycle window drives the rail to $14.7\,\mu$A/bit,
matching the architectural $14.8\,\mu$A/bit, while a large
window of hundreds of cycles collapses it to $\approx\!20\%$ of that. The
recovery curve tracks the architectural gadget (\autoref{fig:dramhw:spectre:cpa}): at $\approx\!60$k 
traces, it recovers $63/64$ bytes in a 2-candidate residual, the same result as the
architectural attack, in $\approx\!18$\,h. The block is recovered while
it is read only on the squashed speculative path, never by a committed
instruction.

\section{Discussion}
\label{sec:disc}
\subsection{Generality of \sysname-RE}
\label{sec:disc:re-generality}
The recovery framework of \autoref{sec:dramre:recovery} is
side-channel-agnostic. It requires only an oracle that distinguishes the level
at which two addresses agree in the DRAM hierarchy; any side channel providing
this instantiates the same procedure and reaches the same sample bound
(\autoref{thm:dramre:bound}). The advantage over a binary same-bank oracle
is structural: it measures each level independently and conditionally,
replacing one exponentially rare collision with a sequence of cheaper
conditional events. The saving compounds with hierarchy depth and depends
only on the hierarchy, not on the side channel, as shown in Appendix~\ref{appendix:genericity}, \autoref{fig:appendix:landscape}.

This generality is bounded by one condition: the side channel must produce a
distinct, separable current level at each hierarchy level being recovered.
We demonstrated this for bank, sub-channel, and chip on the Jetson~TX2,
where the four agreement classes form non-overlapping current peaks
(\autoref{sec:dramre:recovery-results}). Channels, ranks, and bank groups remain untested, since our platforms do not support them. We expect them to separate: a bank group imposes a longer column-to-column
delay ($t_{\mathrm{CCD\_L}}$ vs.\ $t_{\mathrm{CCD\_S}}$), and a rank switch
adds bus turnaround and on-die-termination reconfiguration. Each shifts
command density or I/O current, so each should yield a distinct level.
Confirming this hypothesis on a multi-rank, bank-grouped DDR4/5 platform remains future
work.

\subsection{Applicability of \sysname-DATA}
\label{sec:disc:data-applicability}
\sysname-DATA is gated by one hardware condition: a power rail supplying
the DRAM array core must be readable from unprivileged software. Three
cases follow. Where a core-array rail is exposed, as \texttt{DDR\_VDD2\_A}
is on the Pi~5, both leakage terms are available and the full attack runs.
Where only an I/O rail such as \texttt{DDR\_VDDQ\_A} is exposed, the attack
is limited to the static Hamming-weight channel: I/O rails carry no
Hamming-distance cross-term (\autoref{sec:dramhw:toggle}), so value
recovery is unavailable. Where no DRAM rail reaches userspace, the attack
does not apply; many smartphone SoCs fall here, restricting PMIC telemetry
to privileged software or aggregating domains into a single supply that
hides the array current.

Which platforms qualify is an open question.
The requirement, a core-array rail with unprivileged telemetry, varies across vendors and form factors; constantly surveying its exposure on AMD, Intel, Qualcomm, and embedded Arm platforms is therefore of major importance.


\subsection{Portability}
The data attacks work across boards without retuning. We rerun both
modes, Hamming-weight and value recovery, on a second Pi~5 with 2\,GB of
memory instead of 16\,GB, and a new secret. The smaller capacity moves the
channel-select bit from bit~31 to bit~29; the primitive
handles this with no manual input by choosing the target layout from the
measured per-bit current. The rail is weaker on this board ($\gamma$ about
half the 16\,GB value), so recovery needs more traces. At $120$k traces
($\approx\!34$\,h) the blind value attack recovers $47/64$ bytes at
rank~$1$ and the rest at rank~$\le 5$. The Hamming-weight mode reproduces
just as cleanly, recovering $62/64$ bytes within $\pm 1$ bit of the true
weight (MAE $0.30$ bit).

\subsection{Memory Block Placement Constraint}
\label{sec:disc:loc}
Our gadget places $G$ and $V$ in the same DRAM row
(\autoref{sec:dramhw:primitive}), but this is not required.
\autoref{tab:disc:placement} reports the per-bit swing $\gamma$ and the recovery outcome across placements on the Pi~5.
The only constraint is a shared channel.
A full cross-bank CPA ($G$ and $V$ in different banks of the same channel) recovers the secret at $\rho\approx0.046$, against the same-row $\rho\approx0.050$, and leakage is stronger for memory blocks in different banks of the same channel ($\gamma=78\%$) than for those in the same bank but different rows ($\gamma=56\%$).
Switching to a different channel reduces the attack to $0/64$: $G$ and $V$ no longer share the DQ bus.
In our threat model, the same-channel placement, achievable using \sysname{}-RE's extra-hierarchical information, is a key condition and also easier to satisfy for unprivileged attackers than the same-row condition.

\begin{table}[h]
  \centering
  \small
  \caption{$G$/$V$ placement vs.\ leakage on the Pi~5.}
  \label{tab:disc:placement}
  \begin{tabular}{@{}lcc@{}}
    \toprule
    Placement & $\gamma$ vs.\ same-row & Recovery \\
    \midrule
    Same row                     & $100\%$        & yes (baseline) \\
    Different bank, same channel & $78\%$  & yes \\
    Same bank, different row     & $56\%$         & yes (weakest) \\
    Different channel            & ---            & no ($0/64$) \\
    \bottomrule
  \end{tabular}
\end{table}

\subsection{Threat-model realism}
While \sysname only requires code execution and the existence of commodity sensors, \sysname-DATA requires loading the victim line from the DRAM.
A first possibility to achieve this on interesting lines is to leverage certain kernel ioctl allowing users to load kernel addresses from the userspace~\cite{kogler2023collidepower}.
The second option is to leverage speculative loads through Spectre gadgets as demonstrated in~\autoref{sec:dramhw:spectre}.
Such gadgets have been found in production kernels in previous work~\cite{koruyeh2018spectre,ret2spec}, and the discovery of such gadgets is an orthogonal field of work. 

In our evaluation of \sysname-DATA, we use a kernel module that loads two memory blocks: an attacker-controlled block and a block containing secret data.
Although our evaluation uses a customized workload, similar access patterns can appear in real-world applications.
For example, a service decrypting attacker-supplied payloads using a cryptographic library like OpenSSL~\cite{openssl} may access attacker-controlled ciphertext and secret key.
By varying the payload contents and measuring the resulting power consumption, an attacker could potentially apply \sysname-DATA to recover the accessed secret key.



\subsection{Comparison Against Collide+Power}
\label{sec:disc:collidepower}
\begin{figure}[t]
  \centering
  \begin{subfigure}[t]{0.49\linewidth}
    \centering
    \includegraphics[width=\linewidth]{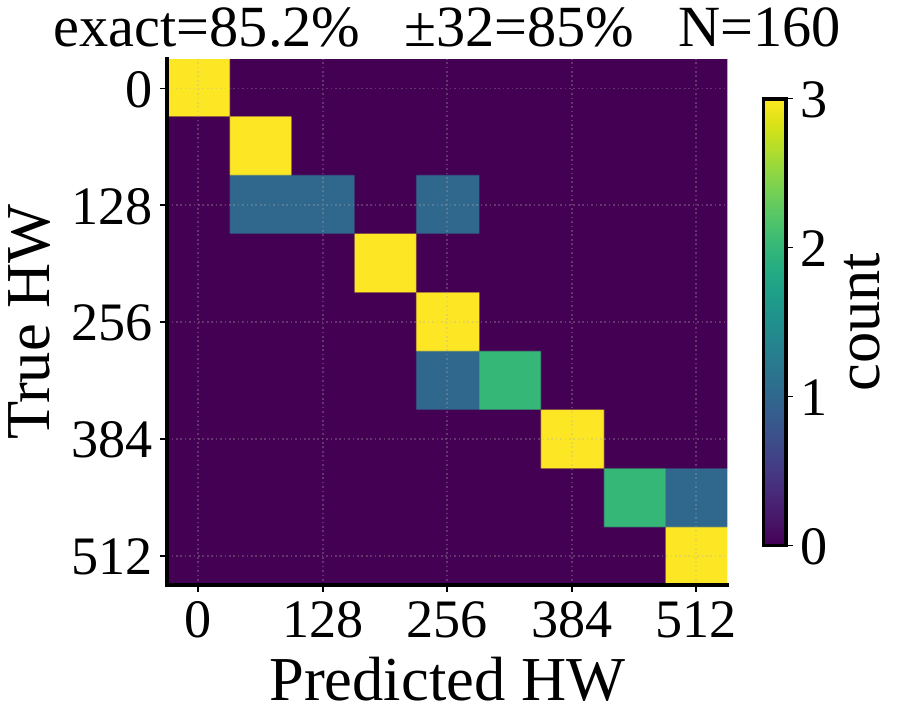}
    \caption{Hamming-weight recovery.}
    \label{fig:dramnc:hw:cm}
  \end{subfigure}
  \begin{subfigure}[t]{0.49\linewidth}
    \centering
    \includegraphics[width=\linewidth]{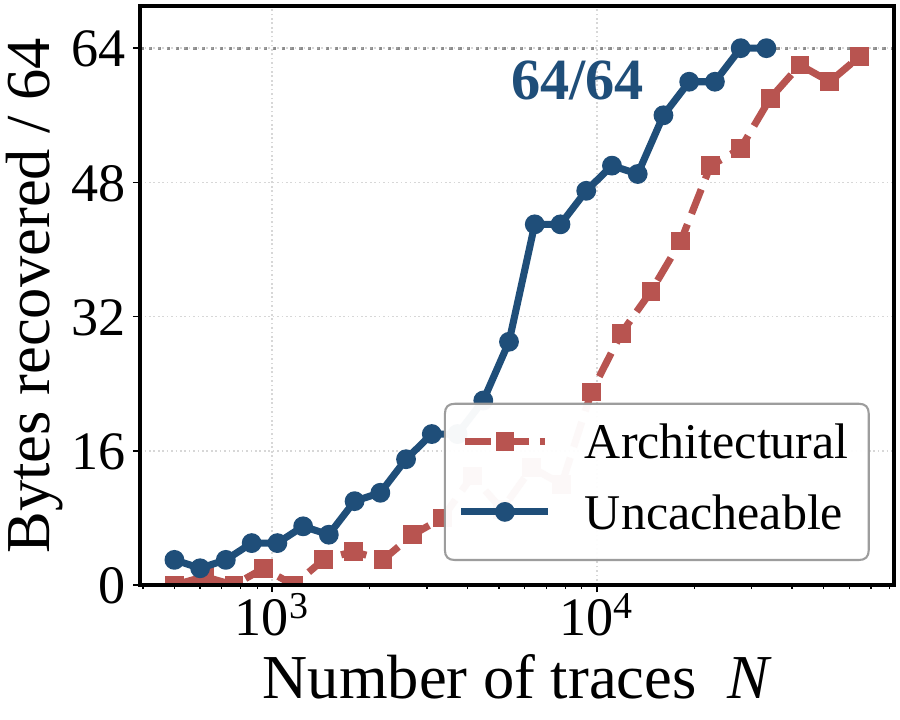}
    \caption{Value recovery with CPA.}
    \label{fig:dramnc:hd:cpa}
  \end{subfigure}
  \caption{\sysname-DATA on \emph{uncacheable} memory.}
  \label{fig:dramnc:hw}
\end{figure}

\sysname leaks data as it moves between DRAM and the controller, so it
reaches any line the victim reads from memory, whereas Collide+Power~\cite{kogler2023collidepower}
leaks the transition between two values inside a cache line and reaches
only cacheable data. To show the difference, we map the gadget blocks as
\emph{Normal non-cacheable} and drop the eviction step; we still observe the leakage. This models a victim that processes attacker data next to a
secret in an uncacheable region, such as DMA descriptors or device-mapped
buffers, where Collide+Power has no signal. For the template attack, the
flush-free gadget profiles the weight of an uncacheable line, and the
joint z-scored template (\autoref{sec:dramhw:hw}) recovers it at $85\%$
exact (\autoref{fig:dramnc:hw:cm}); the CPA recovers all
$64/64$ bytes at $N\approx3.3\times10^4$ traces
(\autoref{fig:dramnc:hd:cpa}), fewer than the cacheable gadget needs for
$63/64$. \sysname thus reaches uncacheable memory at parity with the
cacheable gadget, a victim class Collide+Power cannot reach.

\subsection{Mitigations}
\label{sec:mitigations}
Revoking unprivileged access to PMIC registers eliminates the software
attack path, and it constitutes a primary practical mitigation.
It removes the adversary's access to the leakage rather than the leakage itself, so stronger adversaries can still maintain a measurement point: a privileged attacker operating in the normal world could still perform a PMIC-based attack on the trusted world~\cite{le2025not}.
Furthermore, a root-level adversary, or any attacker with a physical measurement point, retains full access regardless of the access-control policy.
Near-field EM probes above the SoC can recover DRAM bus signals without any software sensor access~\cite{maillard2024emcache}, and embedded circuits in SoC FPGAs have demonstrated the ability to measure DDR activity~\cite{quere2026snoopypower}.

\section{Conclusion}
\label{sec:conclusion}

This work demonstrated that the unprivileged low-resolution, aggregate power sensors of modern Arm SoCs can be used to measure the power signature of DRAM activity.
Building on this, we built two attack families.
\sysname-RE recovers complete DRAM addressing functions from
userspace, succeeds on platforms where the timing channel fails
outright, and exposes more hierarchy information than state-of-the-art methodologies.
\sysname-DATA shows that the same primitive can successfully recover the Hamming weight
of a transferred memory block, or the byte values moving from the DRAM.
These attacks can be mounted directly, or by leveraging a Spectre gadget.
Together, these attacks demonstrate that power sensors are a powerful and important threat to SoC security, even with low timing or power resolution, and completely from unprivileged software.

\section*{Use of Generative AI.}
Generative AI has been used for editorial and light development purposes.
All outputs were reviewed by the authors for accuracy and coherence; originality remains the authors' contribution and responsibility.

\section*{Acknowledgments}
This research is supported in part by the ANR within the framework of the PIA EUR CyberSchool project (ANR-18-EURE-0004) and the European Research Council (ERC project FSSec 101076409).
Additional funding was provided by generous gifts from Intel.
Any opinions, findings, and conclusions or recommendations expressed in this paper are those of the authors and do not necessarily reflect the views of the funding parties.

\bibliographystyle{plain}
\bibliography{shrew}

\appendices
\section{Recovered Bank-Selection Functions}
\label{appendix:mapping}
\autoref{tab:appendix:mapping} lists the bank-selection functions
recovered by \sysname on each evaluation platform. We use the compact
notation of prior DRAM
work~\cite{pessl2016drama,plin2026knockknock}: an entry $[i,\,j]$ denotes
$\mathrm{PA}[i]\oplus\mathrm{PA}[j]$, and a single $[i]$ denotes the direct
selector $\mathrm{PA}[i]$. The bank index concatenates the listed
functions in order.

\begin{table*}[t]
  \centering
  \small
  \caption{Bank-selection functions recovered by \sysname across the three
  evaluation platforms. Jetson functions are reported up to the rank-$2$
  basis of \autoref{sec:dramre:recovery-results}.}
  \label{tab:appendix:mapping}
  \begin{tabular}{@{}l l l@{}}
    \toprule
    \textbf{Platform} & \textbf{Hierarchy} & \textbf{Retrieved Functions} \\
    \midrule
    AMD ZCU102      & 4 banks  & $[13],\ [14]$ \\
    \addlinespace[2pt]
    Raspberry~Pi~5     & 16 banks & $[12,\,33],\ [31,\,32],\ [31],\ [12]$ \\
    \addlinespace[2pt]
    NVIDIA Jetson~TX2
      & 2 chips $\times$ 2 sub-channels $\times$ 4 banks
      & $c{:}\ [13]$;\quad $s{:}\ [12,\,13]$;\quad
        $(f_1, f_2){:}\ \{18,19,21,22,24,25,27,28\},$ \\
    & & \hphantom{$c{:}\ [13]$;\quad $s{:}\ [12,\,13]$;\quad $(f_1, f_2){:}\ $}%
        $\{14,17,18,20,21,23,24,26,27,29\}$ \\
    \bottomrule
  \end{tabular}
\end{table*}

\section{\sysname-RE Genericity and Theoretical Applicability}
\label{appendix:genericity}
\begin{figure}[h!]
  \centering
  \includegraphics[width=\linewidth]{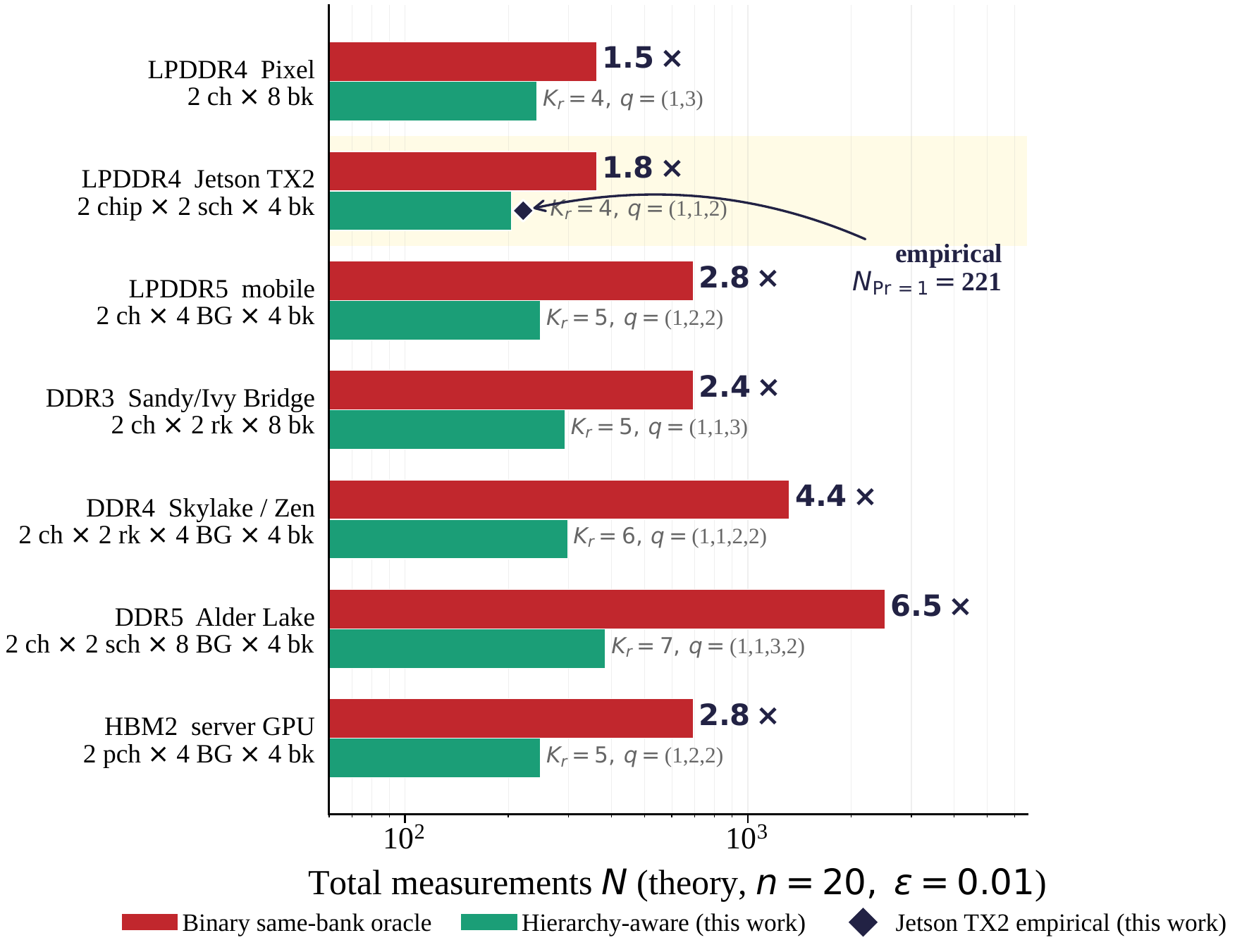}
  \caption{\sysname-RE genericity across hierarchy depth.}
  \label{fig:appendix:landscape}
\end{figure}

\autoref{fig:appendix:landscape} projects \sysname-RE onto the space of
DRAM hierarchies it can address: any controller that exposes $r>1$ levels
of agreement admits the multi-level recovery, and the per-trace saving
grows with depth.

\section{Kernel Gadgets}
\label{sec:appendix:gadgets}
All three gadgets instantiate the data-dependent stimulus
$\mathcal{H}_{\mathrm{D}}(G,V)$: each iteration drives one
$G\!\rightarrow\!V$ transfer over the bus and sinks the loaded data to
defeat dead-code elimination. They differ only in how the transfer reaches
DRAM. In every listing \texttt{x11}$=G$, \texttt{x13}$=V$,
\texttt{x15}$=$sink, and \texttt{x12}$=$iteration count; register binding
and clobbers are omitted.

\prettyparagraph{Reference gadget.}
\autoref{lst:gadget:ref} is the committed gadget used for Hamming-weight
and value recovery (\autoref{sec:dramhw}). \texttt{dc civac} flushes $G$
and $V$ to DRAM and a \texttt{dsb} orders the flush, so the committed
$G\!\rightarrow\!V$ loads each miss the cache and drive one clean bus
transition.

\needspace{16\baselineskip}
\begin{lstlisting}[style=gadget,caption={Reference gadget: committed
$G\!\rightarrow\!V$ with explicit flush.},label={lst:gadget:ref}]
1:  dc   civac, x11          // flush G to DRAM
    dc   civac, x13          // flush V to DRAM
    dsb  ish
    ldp  q0, q1, [x11, #0]   // read G   (committed)
    ldp  q2, q3, [x11, #32]
    ldp  q4, q5, [x13, #0]   // read V   -> G->V bus transition
    ldp  q6, q7, [x13, #32]
    umov x16, v3.d[0]        // consume + sink
    str  x16, [x15]
    umov x16, v7.d[0]
    str  x16, [x15]
    subs x12, x12, #1
    b.ne 1b
\end{lstlisting}

\prettyparagraph{Uncacheable gadget.}
\autoref{lst:gadget:nc} drops the flush and barrier
(\autoref{sec:disc:collidepower}): $G$ and $V$ are mapped \emph{Normal
non-cacheable}, so every load reaches DRAM on its own. The bus transition
is identical to the reference gadget, with no cache maintenance.

\needspace{15\baselineskip}
\begin{lstlisting}[style=gadget,caption={Uncacheable gadget: no flush, no
barrier; the Normal-non-cacheable mapping sends every load to
DRAM.},label={lst:gadget:nc}]
1:  // G,V are Normal non-cacheable: every load reaches DRAM
    ldp  q0, q1, [x11, #0]   // read G
    ldp  q2, q3, [x11, #32]
    ldp  q4, q5, [x13, #0]   // read V   -> G->V bus transition
    ldp  q6, q7, [x13, #32]
    umov x16, v3.d[0]        // consume + sink
    str  x16, [x15]
    umov x16, v7.d[0]
    str  x16, [x15]
    subs x12, x12, #1
    b.ne 1b
\end{lstlisting}

\prettyparagraph{Speculative gadget.}
\autoref{lst:gadget:spec} reads $V$ only on a mispredicted path
(\autoref{sec:dramhw:spectre}). The \texttt{bl} pushes label~\texttt{3}
onto the return-stack buffer, training it to predict \texttt{3} as the
return target. A dependent ALU loop holds the speculation window while the
$G\!\rightarrow\!V$ burst issues; the \texttt{ret} then resolves to
label~\texttt{4}, commits the architectural path, and squashes the burst,
but the linefills it dispatched already complete on the rail.

\needspace{20\baselineskip}
\begin{lstlisting}[style=gadget,caption={Speculative (Spectre-RSB) gadget:
$G\!\rightarrow\!V$ issues only on the squashed return-mispredict
path.},label={lst:gadget:spec}]
    adr  x9, 4f
1:  dc   civac, x11          // flush G to DRAM
    dc   civac, x13          // flush V to DRAM
    dsb  ish
    bl   2f                  // train RSB: push &3f as return target
3:  ldp  q0, q1, [x11, #0]   // SPEC window: read G
    ldp  q2, q3, [x11, #32]
    ldp  q4, q5, [x13, #0]   // spec G->V bus transition
    ldp  q6, q7, [x13, #32]
2:  mov  x6, ak              // dependent ALU delay holds the window
6:  subs x6, x6, #1
    b.ne 6b
    add  x30, x9, x6         // x30 = &4f ; ret resolves after the delay
    ret                      // architecture -> 4f, RSB speculation -> 3f
4:  umov x16, v3.d[0]        // consume + sink
    str  x16, [x15]
    subs x12, x12, #1
    b.ne 1b
\end{lstlisting}

\section*{Ethics Considerations}
All experiments were conducted on hardware owned by the authors, in a
laboratory setting, with no third-party user, production service, or
personal data involved. The secrets recovered in \autoref{sec:dramhw} are
values the authors planted for measurement; no real credential or key was
targeted. Because \sysname exploits a hardware and platform-integration
property rather than a software defect the authors could patch, we
followed a coordinated-disclosure process: we notified the national CERT
on June~09,~2026 and Raspberry~Pi~Ltd on June~10, ahead of publication,
and disclosure to the other affected vendors is proceeding with the
CERT's assistance. We believe publication strengthens defenses on
balance, since the mitigation we recommend in \autoref{sec:mitigations}
is an access-control change that platform vendors and distributions can
apply without hardware modification, and only parties aware of the
exposure can apply it.

\end{document}